\documentclass{iopart}
\usepackage{iopams}
\usepackage{graphicx}
\usepackage{braket}
\usepackage{cite}

\newcommand{\abs}[1]{\ensuremath{\left\vert#1\right\vert}}

\begin{document}

\title[Wave Functions, Quantum Diffusion in Quasiperiodic Tilings]{Wave Functions, Quantum Diffusion, and Scaling Exponents in Golden-Mean Quasiperiodic Tilings}

\author{Stefanie Thiem and Michael Schreiber}
\address{Institut f\"ur Physik, Technische Universit\"at Chemnitz, D-09107 Chemnitz, Germany}
\ead{stefanie.thiem@physik.tu-chemnitz.de}

\begin{abstract}
We study the properties of wave functions and the wave-packet dynamics in quasiperiodic tight-binding models in one, two, and three dimensions. The atoms in the one-dimensional quasiperiodic chains are coupled by weak and strong bonds aligned according to the Fibonacci sequence. The associated $d$-dimensional quasiperiodic tilings are constructed from the direct product of $d$ such chains, which yields either the hypercubic tiling or the labyrinth tiling. This approach allows us to consider rather large systems numerically. We show that the wave functions of the system are multifractal and that their properties can be related to the structure of the system in the regime of strong quasiperiodic modulation by a renormalization group (RG) approach. We also study the dynamics of wave packets to get information about the electronic transport properties. In particular, we investigate the scaling behaviour of the return probability of the wave packet with time. Applying again the RG approach we show that in the regime of strong quasiperiodic modulation the return probability is governed by the underlying quasiperiodic structure. Further, we also discuss lower bounds for the scaling exponent of the width of the wave packet and propose a modified lower bound for the absolute continuous regime.
\end{abstract}


\submitto{\JPCM}
\maketitle

%******************************************************************************************************************************************
\section{Introduction}\label{sec:introduction}
%******************************************************************************************************************************************

In the 1970s works by Penrose and Ammann showed that the Euclidean space can be filled gapless and non-overlapping by two or more tiles which are arranged
in a nonperiodic way according to matching rules \cite{BIMathApp.1974.Penrose}. These quasiperiodic tilings showed a hierarchical structure and long-range oriental order. The study of such tilings was strongly intensified after the discovery of quasicrystals by Shechtman et al.\ in 1982 \cite{PhysRevLett.1984.Shechtman}. These materials are characterized by a perfect long range order without having a three-dimensional translational periodicity. The former is manifested by the occurrence of sharp spots in the diffraction pattern and the latter in the occurrence of rotational symmetries forbidden for conventional crystals \cite{PhysProp.1999.Fujiwara, NoticeAMS.2005.Senechal}. While the diffractions patterns of quasicrystals were not explainable by the structure of crystals with their periodic repetition of unit cells, the mathematical concept of quasiperiodic tilings was suitable to describe the structure of this new material class \cite{PhysRevLett.1984.Levine}.

Various experimental investigations revealed rather exotic physical properties. For instance quasicrystalline surfaces are anti-adhesive in combination with a high level of hardness making them suitable for the production of coatings of cookware, engines, tools, etc. \cite{Patent.1993.Dubois}. Further, they possess a rather low thermal and electrical conductance although they contain a high amount of well-conducting elements \cite{PhysicalProperties.1999.Stadnik, JMathPhys.1997.Roche}. This motivated extensive research to obtain a better theoretical understanding of the structure and the physical properties of these materials. Today several exact results are known for one-dimensional quasicrystals \cite{MathQuasi.2000.Damanik, PhysRevB.1987.Kohmoto, JPhysFrance.1989.Sire}. However, the characteristics in two or three dimensions have been clarified to a much lesser degree and are mainly based on numerical studies. Due to the lack of translation symmetry in quasicrystals, one has to study very large approximants to obtain an insight into the physical properties of macroscopic quasicrystals. In addition many physical properties in quasiperiodic systems show a power-law behaviour with respect to time or system size, which indicates the absence of a typical time/length scale in the system. That behavior is in correspondence with the self-similarity of quasicrystalline structures, what allows scaling between the various hierarchic levels of the tilings. The resulting scaling exponents provide an insight into the properties of the macroscopic systems. Such a scaling approach was also used for the numerical determination of the electronic properties for the Penrose tiling, the Ammann-Kramer-Neri tiling, or octagonal tiling models \cite{Quasicrystals.2003.Grimm, PhysRevB.1992.Passaro}. However, all these calculations were performed for relatively small systems, making it difficult to determine the exact scaling behaviour.

To obtain a better insight into the properties of quantum-mechanical wave functions and the quantum diffusion in quasicrystals, we study here models of $d$-dimensional quasicrystals with a separable Hamiltonian in a tight-binding approach \cite{EurophysLett.1989.Sire}. This method is based on the Fibonacci  sequences, which describes the weak and strong couplings of atoms in a quasiperiodic chain. The Fibonacci chain is the one-dimensional analogue of the Penrose tiling and the icosahedral quasicrystal \cite{PhysRevB.1986.Socolar} and, hence, it is related to quasicrystals with five-fold and ten-fold rotational symmetries. Our higher-dimensional tilings are then constructed as a direct product of these chains. Also the eigenstates of these tilings can be directly calculated by multiplying the wave functions and by adding/multiplying the energies of the quasiperiodic chains, which yields the hypercubic tiling or the labyrinth tiling, respectively. This approach allows us to study very large systems with up to $10^{10}$ sites in three dimensions based on the solutions in one dimension.

The quantum diffusion in the Fibonacci chain and the corresponding hypercubic tiling has been studied quite intensively in the past \cite{PhysRevA.1987.Abe, JPhys.1995.Zhong, JAC.2002.Lifshitz,PhysRevB.2004.Sanchez}. Further, in our previous work we obtained more insight into the electronic eigenstates and the transport properties of the labyrinth tiling. For instance, we showed that the product structure of the labyrinth tiling leads to a connection of the multifractal properties of the wave functions in different dimensions \cite{JPhysCS.2010.Thiem,EPJB.2011.Thiem}. In recent years also some new properties for one-dimensional quasiperiodic chains were found. These include the step-like behaviour of the wave-packet dynamics of metallic-mean chains reported by us \cite{PhysRevB.2009.Thiem} and the discovery of log-time-periodic oscillations of wave packets in the Fibonacci chain \cite{PhilMag.2011.Lifshitz}.

In this paper we report new results for the multifractal properties of wave functions in quasiperiodic systems and for the return probability of the wave packets. In particular, we study the participation numbers which provide an insight into the distribution of the probability density of electronic wave functions and allow us to distinguish extended and localized states. This can be related to the electronic transport properties, because e.g.\ Bloch functions indicate ballistic transport and localized wave functions are associated with insulators. To describe the electronic transport properties more quantitatively two quantities are usually studied: the mean square displacement and the temporal autocorrelation function. The scaling exponent $\beta$ of the mean square displacement is connected to the electrical conductivity of a quasicrystal according to the generalized Drude equation \cite{PhysRevLett.1997.Schulz-Baldes}. In our recent work we showed that the scaling exponent $\beta$ can be determined analytically for the labyrinth tiling in the regime of strong quasiperiodic modulation by a renormalization group (RG) approach \cite{PhysRevB.2012.Thiem}. In the present paper we focus on the temporal autocorrelation function, which is directly related to the properties of the (local) density of states. Further, it describes the decay of the wave packet which can be linked to the quantum diffusion of electrons in the system. We also show that for systems with absolute continuous energy spectra an improved lower bound for the exponent $\beta$ can be obtained, which is not restricted to the systems studied here.

The paper is structured in the following way. In Sec.\ \ref{sec:state-of-art} we briefly describe the construction rules for the $d$-dimensional separable tilings. We also introduce the participation numbers and the temporal autocorrelation function and give an overview of known results for these quantities in quasiperiodic systems. In Section \ref{sec:rg-approach} we briefly describe the RG approach, which is applied in the following two sections. This is followed by the presentation of new results for the properties of the wave functions in these systems in Sec.\ \ref{sec:waveFunctions}. We also apply the RG approach to show that the wave functions do not become localized even for very small coupling strengths. In Sec.\ \ref{sec:returnProb} we study the wave-packet dynamics with a special focus on the return probability with numerical methods and with the RG approach. Further, the lower bounds for the scaling exponent $\beta$ of the mean square displacement are discussed in Sec.\ \ref{subsec:lower-bound}, and the results are briefly summarized in Sec.\ \ref{sec:conclusion}.

%******************************************************************************************************************************************
\section{Construction, Electronic Structure and Wave-Packet Dynamics of Tiling Models}
\label{sec:state-of-art}
%******************************************************************************************************************************************

\subsection{Separable Tilings with Golden Mean}
\label{sec:model}

The construction of the separable quasiperiodic tilings is based on the Fibonacci chain defined by the inflation rule
$ \mathcal{P} = \{ s \longrightarrow w, w \longrightarrow ws \}$. Starting with the symbol $s$ we obtain after $a$ iterations the $a$th order approximant $\mathcal{C}_a$ of the Fibonacci chain. The length $f_a$ of an approximant $\mathcal{C}_a$ is given by the recursive rule $f_a = f_{a-1} + f_{a-2}$ with $f_0 = f_1 = 1$. Further, the ratio of the lengths of two successive approximants as well as the ratio of the numbers $\#$ of occurrence of the symbols $w$ and $s$ in an approximant approach the golden mean $\tau$ for $a \to \infty$. Thus, we obtain with the continued fraction representation $\tau = [\bar{1}] = [1,1,1,...]$ the relations
\begin{equation}
\label{equ:tau}
\lim_{a \to \infty} \frac{f_a}{f_{a-1}} = \lim_{a \to \infty} \frac{\#_w\left( \mathcal{C}_a \right) } {\#_s\left( \mathcal{C}_a \right)} =  \tau \;.
\end{equation}

Our model describes an electron hopping from one vertex of a quasiperiodic chain to a neighboring one. The aperiodicity is given by the underlying quasiperiodic sequence of couplings, where the symbols $w$ and $s$ denote the weak and strong bonds of the chain. Solving the corresponding time-independent Schr\"odinger equation
\begin{equation}
 \label{equ:octonacci.8}
 \mathbf{H} \ket{\Psi^i}  =  E^i \ket{\Psi^i} \Leftrightarrow t_{l-1,l} \Psi_{l-1}^i + t_{l,l+1} \Psi_{l+1}^{i} =  E^i \Psi_l^i
 \end{equation}
for the Fibonacci chain with zero on-site potentials, we obtain discrete energy values $E^i$ and wave functions $\ket{\Psi^i} = \sum_{l=1}^{f_a+1} \Psi_l^i \ket{l}$ represented in the orthogonal basis states $\ket{l}$ associated to a vertex $l$. If not stated differently we use free boundary conditions, which yields $N_a = f_a + 1$ eigenstates. The hopping strength $t$ in the Schr\"odinger equation is given by the Fibonacci sequence $\mathcal{C}_a$ with $t_{s} = s$ for a strong bond and $t_{w} = w$ for a weak bond ($0 < w \le s$).
For $w=0$ only isolated clusters of single vertices or two strongly coupled vertices occur, and for $w = s$ the system is periodic.

\begin{figure}[t!]
 \centering
 \includegraphics[width=4cm]{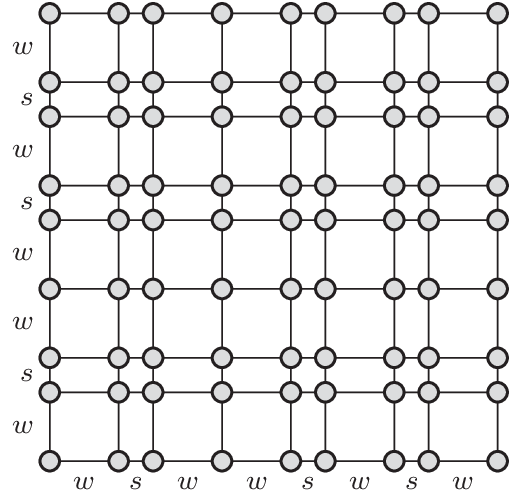}
 \includegraphics[width=4cm]{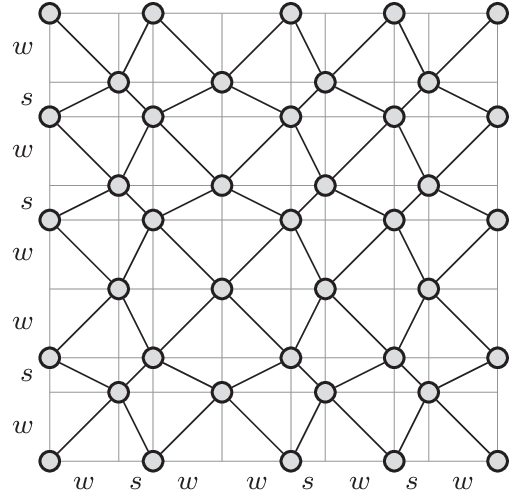}
 \caption{Two-dimensional square tiling $\mathcal{H}_5^{2{\rm d}}$ (left) and labyrinth tiling $\mathcal{L}_5^{2{\rm d}}$ (right) constructed from two Fibonacci chains $\mathcal{C}_5$.}
  \label{fig:tilings}
\end{figure}

The $d$-dimensional separable quasiperiodic tilings are then constructed from the product of $d$ quasiperiodic chains which are perpendicular to each other. For this setup two special cases are known for which the systems become separable \cite{EurophysLett.1989.Sire}:

\emph{Hypercubic Tiling $\mathcal{H}_a^{d{\rm d}}$ } --- This tiling corresponds to the usual Euclidean product of $d$ linear quasiperiodic chains, i.e., only vertices connected by vertical and horizontal bonds interact as shown in Figure \ref{fig:tilings}. Such a tiling is referred to as a square tiling in two dimensions and in general as a hypercubic tiling \cite{Ferro.2004.Ilan, PhilMag.2008.Mandel}. In literature this model has been studied mainly for the Fibonacci sequence \cite{PhysRevB.1990.Ashraff, PhilMag.2008.Mandel, Ferro.2004.Ilan, JAC.2002.Lifshitz, JPhys.1995.Zhong}, and for reasons of comparison we will also present a few results of this model.

\emph{Labyrinth Tiling $\mathcal{L}_a^{d{\rm d}}$} --- In the other case only coupling terms to neighbours along the diagonal bonds are considered (cf.\ Figure \ref{fig:tilings}), where the bond strengths of this tiling equal the product of the corresponding bond strengths of the one-dimensional chains. The resulting grid is denoted as labyrinth tiling.

The Hamiltonians of these tilings are separable. Thus, the eigenstates of the tilings in $d$ dimensions can be constructed from the eigenstates of $d$ one-dimensional chains. In both cases the wave functions of the higher-dimensional tiling are constructed as the product of the one-dimensional wave functions, i.e.,
\begin{equation}
    \Phi_\mathbf{r}^\mathbf{s} = \Phi_{l,m,\ldots, n}^{i,j,\ldots, k} \propto \Psi_{l}^{1i} \Psi_{m}^{2j} \ldots \Psi_{n}^{dk} \;.
\end{equation}
The superscripts $\mathbf{s} = (i,j,\ldots,k)$ enumerate the eigenvalues $E$, and the subscript $\mathbf{r}=(l,m,\ldots,n)$ represents the coordinates of the vertices in the quasiperiodic tiling. However, the energies are treated in a different way for the two cases. In $d$ dimensions the eigenenergies are given by
\begin{equation}
     \label{equ:energies-hypercubic}
    E^\mathbf{s} = E^{i,j,\ldots ,k} = E^{1i} + E^{2j} + \ldots +   E^{dk}
\end{equation}
for the hypercubic tiling and by
\begin{equation}
     \label{equ:energies-labyrinth}
    E^\mathbf{s} = E^{i,j,\ldots ,k} = E^{1i} E^{2j}\ldots  E^{dk}
\end{equation}
for the labyrinth tiling. While for the hypercubic tiling all combinations of the indices $i, j, \ldots, k$ are allowed, for the labyrinth tiling only certain sets yield the correct solution due to symmetries of the eigenfunctions \cite{PhysRevB.2000.Yuan, JPhysCS.2010.Thiem}.

\subsection{Energy Spectra}

For the one-dimensional quasiperiodic systems the energy spectrum is known to be singular continuous \cite{ComMathPhys.2011.Damanik, AMS.2009.Damanik, JStatPhys.1989.Suto}. The properties of the energy spectrum for the hypercubic tiling and the labyrinth tiling can be obtained from studying the scaling behaviour of the width of the energy bands \cite{PhilMag.2008.Mandel, EurophysLett.1989.Sire}. The numerical results for both tilings provided evidence for a transition from a singular continuous to an absolutely continuous energy spectrum upon increasing the coupling parameter $w$ including an intermediate region where both parts coexist \cite{EurophysLett.1989.Sire, PhilMag.2008.Mandel}. The transition parameters depend on the dimension and the underlying quasiperiodic structure. In particular, we like to point out the threshold values $w_{\rm th}^\mathrm{2d} \approx 0.6 s$ in two dimensions and $w_\mathrm{th}^\mathrm{3d} \approx 0.45 s$ in three dimensions above which the spectrum of the hypercubic tiling and the labyrinth tiling becomes absolutely continuous \cite{EurophysLett.1989.Sire, PhilMag.2008.Mandel, PhysRevB.2000.Yuan}.
These transition parameters $w_{\rm th}$ seem to correspond to the threshold values $w_{\rm DOS}$ for which the density of states becomes gapless in two and three dimensions \cite{PhysRevB.2005.Cerovski}.
For $w=0$ the isolated clusters lead to a pure point spectrum and the periodic systems for $w=s$ possess an absolutely continuous spectrum regardless of the dimension.

Further, it was shown analytically that the energy spectrum of the square tiling is singular continuous in the regime of strong quasiperiodic modulation \cite{AMS.2009.Damanik}. Another interesting property is that the energy spectrum of the Fibonacci chain splits into (sub)bands, which can be described by an RG approach proposed by Niu and Nori \cite{PhysRevLett.1986.Niu,PhysRevB.1990.Niu}. We use this property later to derive a relation between the structure and the  properties of the electronic wave function.

%------------------------------------------------------------------------------------------------------------------------------------------
\subsection{Participation Ratio and Multifractality of Wave Functions}
\label{subsec:def-part-ratio}
%------------------------------------------------------------------------------------------------------------------------------------------

In Sec.\ \ref{sec:waveFunctions} we study the multifractal properties of the wave functions in Fibonacci chains and the labyrinth tiling. This includes the computation of the participation ratio $p$, which reflects the fraction of the total number of sites for which the probability measure of the wave function $|\Phi_{\mathbf{r}}^\mathbf{s}|^{2}$ is significantly different from zero. Hence, it provides information about the degree of localization of the wave function $\Phi^\mathbf{s}$ \cite{Quasicrystals.2003.Grimm, ZPhysB.1980.Wegner}. For the discrete positions $\mathbf{r}$ the participation ratio is defined via the inverse participation number $P(\Phi^\mathbf{s})$ of a wave function $\Phi^\mathbf{s}$ and the number of sites $V_a$ in $d$ dimensions as
\begin{equation}
 \label{equ:participation.1}
  p(V_a) =  \left\langle \frac{P(\Phi^\mathbf{s})}{V_a} \right\rangle =   \left\langle \frac{1}{V_a} \left[ \sum_{\mathbf{r}} |\Phi_{\mathbf{r}}^\mathbf{s}|^{4} \right]^{-1} \right\rangle \propto V_a^{-\gamma}\;.
\end{equation}
The expression $\langle \ldots \rangle$ denotes the arithmetic average over all eigenstates in the spectrum or over a certain energy interval. In quasicrystalline systems the participation ratio $p(\Phi,V_a)$ of a state $\Phi$ was found to scale with a power law with respect to the number of sites $V_a$, where the exponent $\gamma(\Phi)$ is connected to the nature of this eigenstate \cite{Quasicrystals.2003.Grimm}.
The wave function of a localized state is characterized by a scaling exponent $\gamma = 1$, and $\gamma = 0$ corresponds to an extended state. Intermediate values of $\gamma$ ($0 < \gamma < 1$) indicate fractal eigenstates, which are neither extended over the whole system nor completely localized at a certain position and show self-similar patterns \cite{PhysRevB.1987.Kohmoto, PhysRevB.1998.Repetowicz}.
The participation ratios within a small energy interval usually fluctuate over a certain range, but the participation ratios and the inverse participation numbers as well as the corresponding scaling exponents do not show any significant trend over the whole energy spectrum for the systems considered here \cite{EPJB.2011.Thiem}.

While the eigenstates of purely periodic systems can be characterized by a single scaling exponent $D = D_q$, the wave functions of chaotic systems and quasicrystals are usually multifractals \cite{APhys.1996.Brandes, PhysRevLett.1994.Guarneri, PhysRevB.1997.Huckestein, IntJPhysB.1995.Salejda2}. The generalized dimensions $D_q$ describe the nature of the multifractality of the eigenstates, where $D_q$ is a monotonically decreasing function of $q$. The dimension $D_0$ is equal to the capacity dimension, which can be understood as the box counting dimension here, $D_1$ is equal to the information dimension, and $D_2$ to the correlation dimension.

The generalized dimensions can be determined either via the generalized inverse participation numbers or via a multifractal analysis. For the first method one investigates the scaling behaviour of the generalized inverse participation numbers averaged over all wave functions of a system with the linear system size $N_a$, i.e.,
\begin{equation}
    \label{equ:gIPN}
    Z_q = \left\langle \sum_\mathbf{r}  \left| \Phi_\mathbf{r}^\mathbf{s} \right|^{2q}  \right\rangle \propto N_a^{-D_q(q-1)} \;.
\end{equation}
However, in Sec.\ \ref{subsec:dq-num} we present numerical results obtained via a multifractal analysis, which shows less numerical problems for negative parameters $q$ in comparison to the generalized inverse participation numbers. The multifractal analysis is based on a standard box-counting algorithm, i.e., in $d$ dimensions the system of linear system size $N_a$ is divided into $B = N_a^d / L^d$ boxes of linear size $L$ \cite{PhysStatSol.1990.Boettger, JPhysA.1986.Castellani, PhysRevLett.1989.Chabra, ChemPhys.1993.Grussbach,PhysRevLett.1991.Schreiber}. For each box $b$ we measure the probability to find the electron in this particular box, i.e., we study the spatial distribution of the wave functions. Hence, for an eigenstate $\Phi_\mathbf{r}^\mathbf{s}$ we determine the probability
\begin{equation}
    \mu_b(\Phi^\mathbf{s}, L) = \sum_{\mathbf{r} \in \, {\rm box}\, b} |\Phi_\mathbf{r}^\mathbf{s} |^2
\end{equation}
and then compute the mass exponent
\begin{equation}
    \tau_q = \lim_{\varepsilon \to 0} \frac{ \langle \ln P(q,\Phi^\mathbf{s},L) \rangle} {\ln \varepsilon}
    =  \lim_{\varepsilon \to 0}\frac{ \langle \ln \sum_{b=1}^{B} \mu_b^q(\Phi^\mathbf{s}, L) \rangle }{\ln \varepsilon}
\end{equation}
by a least squares fit of the quantity $\langle \ln P(q,\Phi^\mathbf{s},L) \rangle$ versus $\ln \varepsilon$ considering box sizes from $L = 10, \ldots, N_a/2$ \cite{PhysRevLett.1989.Chabra, PhysRevLett.1991.Schreiber,PhysRevB.2008.Vasquez}. The quantity $\varepsilon = L/N_a$ denotes the ratio of the box size $L$ and the system size $N_a$. We only consider boxes with integer fractions of $N_a/L$ \cite{PhysRevB.2008.Rodriguez3}. In principle it is possible to improve the quality of the results by adapting the multifractal analysis to work also for non-integer fractions $N_a/L$ and by averaging over different initial positions of the boxes \cite{PhysRevLett.1991.Schreiber,ChemPhys.1993.Grussbach}. However, we do not apply this improved method because it requires significantly more computing time. We also found that the simple partitioning scheme yields sufficiently accurate results for the determination of the mass exponents. The generalized dimensions $D_q$ are easily obtained from the mass exponents according to the relation $D_q = \tau_q / (q-1)$.

%------------------------------------------------------------------------------------------------------------------------------------------
\subsection{Return Probability}
\label{subsec:def-return-prob}
%------------------------------------------------------------------------------------------------------------------------------------------

In the second part of the paper we investigate the dynamics of the quasiperiodic systems by means of the time evolution of a wave packet $\ket{\Upsilon(\mathbf{r}_0,t)} = \sum_{\mathbf{r} \in \mathcal{L}} \Upsilon_\mathbf{r}(\mathbf{r}_0,t) \ket{\mathbf{r}}$, which is constructed from the solutions $\Phi_\mathbf{r}^\mathbf{s}(t) = \Phi_\mathbf{r}^\mathbf{s} e^{-\rmi E^\mathbf{s}t}$ of the time-dependent Schr\"odinger equation obtained by the separation approach. The wave packet is then represented as a superposition of these orthonormal eigenstates $\Phi_\mathbf{r}^\mathbf{s}(t)$ \cite{JPhys.1995.Zhong}, where we assume that it is initially localized at the position $\mathbf{r}_0$ of the quasiperiodic tiling, i.e., $\Upsilon_\mathbf{r} (\mathbf{r}_0,t=0) = \delta_{\mathbf{r}\mathbf{r}_0}$. Hence, with the completeness relation and normalized basis states the wave packet is defined by
\begin{equation}
 \label{equ:wave-pacekt}
 \Upsilon_{\mathbf{r}}(\mathbf{r}_0,t)
  = \sum_{\mathbf{s}} \Phi_{\mathbf{r}_0}^{\mathbf{s}} \Phi_{\mathbf{r}}^{\mathbf{s}}  e^{-\rmi E^{\mathbf{s}} t} \;.
\end{equation}

For a wave packet $\Upsilon_{\mathbf{r}}(\mathbf{r}_0,t)$ initially localized at the position $\mathbf{r}_0$ the temporal autocorrelation function \cite{PhysRevLett.1992.Ketzmerick} $C(\mathbf{r}_0,t)$ corresponds to the integrated probability to be at the position $\mathbf{r}_0$ up to time $t > 0$, i.e.,
 \begin{equation}
   \label{equ:autocorrelation}
    C(\mathbf{r}_0,t) = \frac{1}{t} \int_0^t |\Upsilon_{\mathbf{r}_0}(\mathbf{r}_0,t^\prime)|^2 \rmd t^{\prime} \;.
 \end{equation}
Here, we denote the integrand as return probability $P(\mathbf{r}_0,t)$, i.e.,
 \begin{equation}
 \label{equ:return-prob}
   P(\mathbf{r}_0,t)  = |\Upsilon_{\mathbf{r}_0}(\mathbf{r}_0,t)|^2 \;.
 \end{equation}

The autocorrelation function for an infinite system is expected to decay with a power law $C(\mathbf{r}_0,t) \propto t^{-\delta(\mathbf{r}_0)}$ for $t \rightarrow \infty$ \cite{PhysRevLett.1992.Ketzmerick}, where the scaling exponent $\delta(\mathbf{r}_0)$ depends on the initial position $\mathbf{r}_0$ of the wave packet. Further, it is known that the scaling exponent $\delta(\mathbf{r}_0)$ is equivalent to the correlation dimension $D_2^\mu$ of the local density of states $\varrho(\mathbf{r}_0,E)$ of a system \cite{DukeMathJ.2001.Barbaroux,  JStatPhys.1997.Guerin, CMathPhys.1994.Holschneider, PhysRevLett.1992.Ketzmerick}.
Due to this equivalence we can distinguish two limit cases and their associated spectral properties. For $\delta \rightarrow 0$ the autocorrelation function and the return probability remain constant. A pure point spectrum implies $\delta \rightarrow 0$, but the opposite statement may not be true \cite{PhysRevB.2005.Cerovski, PhysRevLett.1995.delRio}. In one dimension $\delta \rightarrow 1$ corresponds to the ballistic motion of an electron, where the electron moves freely on the chain without scattering \cite{JPhys.1995.Zhong}. In two and three dimensions $\delta \rightarrow 1$ usually does not indicate ballistic transport. Further, an absolutely continuous spectrum implies $\delta \rightarrow 1$ but again the reverse is not necessarily true \cite{PhysRevLett.1995.delRio}.
For quasiperiodic structures one often obtains intermediate values of $\delta$, i.e., $0 < \delta < 1$, which is related to the singular continuous energy spectrum \cite{MathQuasi.2000.Damanik, PhysRevB.1987.Kohmoto, JPhysFrance.1989.Sire, JPhys.1995.Zhong}.

To provide a better insight into the dynamical properties and to compare the properties of different systems, we average the results over different initial positions of the wave packet, i.e., we investigate the scaling behaviour of $C(t) = \langle C(\mathbf{r}_0,t) \rangle \propto t^{-\delta}$.
Previous studies mostly use non-averaged data, which do not adequately display the macroscopic transport properties \cite{PhysRevB.2000.Yuan,JPhys.1995.Zhong}. Due to restrictions in computing time we cannot consider all possible initial positions $\mathbf{r}_0$. In particular, for the Fibonacci chain we have chosen the 500 positions nearest to the center of the chain and in two and three dimensions we have randomly selected 100 vertices in a square of linear size 25 and a cube of linear size 15 around the center of the tilings.

Zhong and Mosseri already studied the temporal autocorrelation function and the return probability for the square and cubic Fibonacci tiling \cite{JPhys.1995.Zhong}. As for this model the higher-dimensional energy values are the sum of the one-dimensional energy values (cf.\ Eq.\ \eref{equ:energies-hypercubic}), the time-evolution operator becomes separable. Therefore, the dynamics of the wave packets in higher dimensions are much better understood than for the labyrinth tiling. They were able to show that
\begin{equation}
 \label{equ:deltaprime-nD}
 \delta_{\rm 1d}^\prime = \delta_{d{\rm d}}^\prime/d
\end{equation}
and that the scaling behaviour of the temporal autocorrelation function is given by \cite{JPhys.1995.Zhong}
\begin{equation}
 \label{equ:delta-nD}
    C(t) \propto
    \cases{ t^{-d\delta_{\rm 1d}} \quad\;\, d\delta_{\rm 1d} < 1 \\
      1 \qquad\quad d\delta_{\rm 1d} \ge 1}
\end{equation}
Further, they found that for the limit case $w = s$ one always obtains $\delta_{\rm 1d}^\prime = 1$. With the same arguments, also the threshold values for the transition to an absolutely continuous energy spectrum can be directly obtained from the one-dimensional scaling exponents $\delta_{\rm 1d}$ according to $d \delta_{\rm 1d} (w_{\rm th}^{d{\rm d}}) \stackrel{!}{=} 1$.

%******************************************************************************************************************************************
\section{RG Approach}
\label{sec:rg-approach}
%******************************************************************************************************************************************

In the regime of strong quasiperiodic modulation it is possible to relate the energy spectra and the wave functions of different approximants of the Fibonacci chain by an RG approach proposed by Niu and Nori \cite{PhysRevLett.1986.Niu,PhysRevLett.1995.Piechon}. We use this approach to show in the following two sections that the wave functions do not become localized even for $w \to 0$. In Sec.\ \ref{sec:returnProb} we also apply the RG approach to find a relation between the quasiperiodic structure of the Fibonacci chain and its return probability. The general idea is that a given system can be transformed (renormalized) by the application of an RG step in such a way that we obtain a new system with a reduced number of degrees of freedom (i.e. sites/bonds). In particular, for the Fibonacci chain for $w = 0$ the isolated sites (atoms) and biatomic clusters (molecules) coupled by a strong bond are visualized in Figure \ref{fig:rg-fib}. This yields three highly degenerate energy levels: $E = 0$ for the atomic sites and $E = \pm s$ for the bonding and antibonding states of the molecules \cite{PhysRevLett.1986.Niu}. For non-zero coupling parameters $w$ one finds a coupling between these isolated atoms and molecules, where the dominant contribution occurs between sites of the same type. Hence, for the Fibonacci chain $\mathcal{C}$ one can distinguish two possible RGs \cite{PhysRevLett.1986.Niu}:

\begin{figure}[b!]
 \centering
 \footnotesize
 \includegraphics[width=8cm]{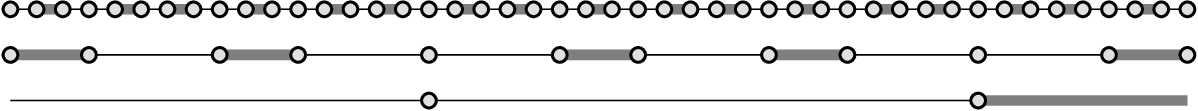}\vspace{0.5cm}
 \includegraphics[width=8cm]{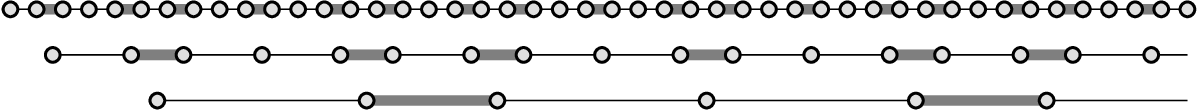}
  \caption{Atomic RG (top) and molecular RG (bottom) for the Fibonacci chain $\mathcal{C}$: in the atomic RG all atomic sites survive an RG step, and in the molecular RG all molecular sites are replaced by a new site. The thick (thin) lines denote the strong (weak) bonds.}
 \label{fig:rg-fib}
\end{figure}

\emph{Atomic RG} --- The renormalized grid after one RG step consists of the atomic sites of the original chain as shown in Figure \ref{fig:rg-fib}. These new sites are then connected by new strong and weak bonds. In particular, atoms separated by one molecule in the original chain become connected by a new strong bond and atoms separated by two molecules in the original chain get connected by a new weak bond.

\emph{Molecular RG} --- In one RG step all molecular sites of the original chain are replaced by new atomic sites (cf.\ Figure \ref{fig:rg-fib}). The atoms in the renormalized grid are connected by new strong bonds between the sites of originally neighboring molecules and by new weak bonds between the sites for molecules that were separated by an atomic site.

For both RG approaches we obtain a new Fibonacci chain, which is scaled in length and energy. The derivation of the length scalings $c$ and the scaling factors $z$ for the new bond strengths of an approximant under the application of one RG step is based on the Brillouin-Wigner perturbation theory. For the Fibonacci chain $\mathcal{C}$ one obtains for the \cite{PhysRevLett.1986.Niu, PhysRevA.1987.Abe}
\begin{itemize}
 \item \emph{atomic RG} a length scaling of $c = \tau^{-3}$ and an energy scaling of $\bar{z} = w^2/s^2$ and for the\vspace{-0.2cm}
 \item \emph{molecular RG} a length scaling of $c = \tau^{-2}$ and an energy scaling of $z = w/2s$.
\end{itemize}

In other words, the energy spectrum $\mathcal{E}_a(s,w)$ of the Hamiltonian $\mathbf{H}_a(s,w)$ of the Fibonacci chain $\mathcal{C}_a$ can be related to the spectrum of three Hamiltonians of smaller approximants \cite{PhysRevLett.1995.Piechon}. With the scaling relations given above, it can be shown that the energy spectrum splits according to \cite{PhysRevLett.1995.Piechon}
\begin{equation}
  \label{equ:enegry-spectrum-fib}
  \mathcal{E}_a(s,w) \longrightarrow
  \cases{
   -s + z \mathcal{E}_{a-2}(s,w) \\
   \bar{z} \mathcal{E}_{a-3}(s,w) \\
   +s + z \mathcal{E}_{a-2}(s,w) } \quad.
\end{equation}
This means that the central main band of the energy spectrum corresponds to the spectrum $\mathcal{E}_{a-3}(s,w)$ of the Hamiltonian $\mathbf{H}_{a-3}(\bar{z}s, \bar{z}w)$ comprising the $f_{a-3}$ atomic sites for $w = 0$ scaled by a factor $\bar{z}$. The left and right main bands at energies $\pm s + z \mathcal{E}_{a-2}(s,w)$ correspond to the bonding and antibonding states of the spectra of the Hamiltonian $\mathbf{H}_{a-2}(z s, z w)$ for the $f_{a-2}$ molecular sites \cite{PhysRevLett.1996.Piechon}.

\begin{figure}[t!]
  \centering
  \includegraphics[width=8cm]{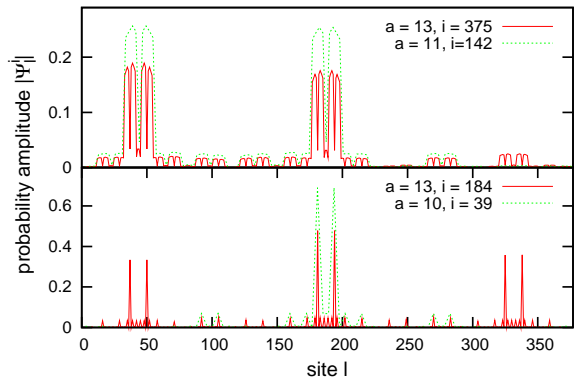}
  \caption{Scaling of the amplitudes of the wave functions $\Psi_l^i$ for the Fibonacci chain $\mathcal{C}_a$ for the molecular RG (top) and the atomic RG (bottom) for $w=0.1$ and $s=1$.}
  \label{fig:wave-functions-rg}
\end{figure}

The recursive structure of the chains also has an influence on the properties of the individual wave functions. In particular, for the Fibonacci chain $\mathcal{C}_a$ with periodic boundary conditions Pi\'echon provided a quantitative result for the amplitudes of the wave functions for different approximants $a$ \cite{PhysRevLett.1996.Piechon}. Note that in this case the number of sites/eigenstates is $N_a = f_a$. For the energy levels $E^{i}$ ($f_{a-2} < i + f_{a-2} \le f_{a-1}$) in the central main band of the energy spectrum associated to the atomic sites, the relation is given by
\begin{equation}
 \label{equ:wave-functions-atomRG}
 |\Psi_{l,a}^{i+f_{a-2}}(E)| \simeq \sqrt[4]{\tau^{-3}} | \Psi_{l^\prime,a-3}^i(E/\bar{z})| \;.
\end{equation}
Likewise, one obtains the recursive relations for wave functions associated to the energy levels $E^i$ belonging to the left main band ($1 \le i \le f_{a-2}$) and to the right main band ($f_{a-1} < i + f_{a-1}\le f_a$) of the energy spectrum:
%\numparts
\begin{equation}
 \label{equ:wave-functions-molRG}
 \eqalign{
 \Psi_{l^+,a}^i(E) - \Psi_{l^-,a}^i(E)   &\simeq \frac{2}{\sqrt{\tau}} \Psi_{l^\prime,a-2}^i \left(\frac{E + s}{z}\right) \cr
 \Psi_{l^+,a}^{i+f_{a-1}}(E) + \Psi_{l^-,a}^{i+f_{a-1}}(E)  &\simeq \frac{2}{\sqrt{\tau}} \Psi_{l^\prime,a-2}^i \left( \frac{E - s}{z} \right) .}
\end{equation}
%\endnumparts
The sites $l^+$ and $l^-$ are those linked by a bond of strength $s$, which is replaced by the new site $l^\prime$ during one step of the molecular RG. These are approximate results because the atomic (molecular) RG approach ignores the probability of being on a molecular (atomic) site. Comparing this with the numerical results in Figure \ref{fig:wave-functions-rg}, we see that this nevertheless is a good approximation since the probability density on the other type of cluster is much lower in the regime of strong quasiperiodic modulation ($w\ll s$).

Note that this RG approach is different from the RG approach used by Wang and Sanchez for the analysis of the electronic transport in Fibonacci chains and the corresponding hypercubic tilings \cite{PhysRevB.2001.Sanchez, PhysRevB.2004.Sanchez}. With their renormalization rules they were able to derive an iterative procedure for the evaluation of Kubo-Greenwood formula at zero temperature. While this method allows one to efficiently compute the conductivity for large quasiperiodic systems, the RG approach applied in this paper allows us to directly connect the structure of the systems to certain properties of the wave functions and the transport properties.

%******************************************************************************************************************************************
\section{Multifractality of Wave Functions}
\label{sec:waveFunctions}
%******************************************************************************************************************************************

In this section we first present numerical results for the multifractal properties of the wave functions for the Fibonacci systems. We then show that the multifractality is directly related to the structure of the quasiperiodic systems for small coupling strengths $w$ and that even for $w \to 0$ the wave functions remain multifractals.

%------------------------------------------------------------------------------------------------------------------------------------------
\subsection{Numerical Results for Multifractality of Wave Functions}
\label{subsec:dq-num}
%------------------------------------------------------------------------------------------------------------------------------------------

\begin{figure}[t!]
  \centering
  \includegraphics[width=8cm]{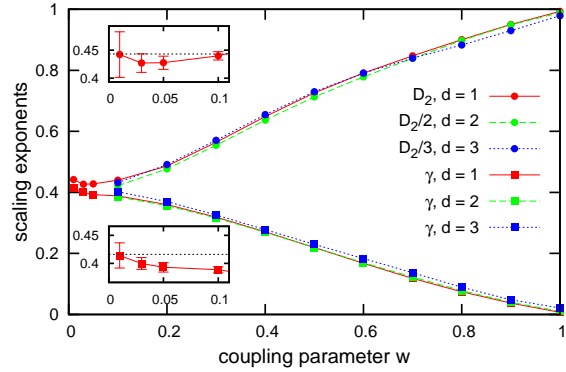}
  \caption{Correlation dimension $D_2$ of the inverse participation number $Z_2$ and the scaling exponent $\gamma$ of the participation ratio $p$ averaged over all wave functions of the systems for $s=1$. The largest considered systems are the Fibonacci chain $\mathcal{C}_{20}$ with $10\,947$ sites and the labyrinth tilings $\mathcal{L}_{17}^{\rm 2d}$ with $3\,335\,945$ sites and $\mathcal{L}_{12}^{\rm 3d}$ with $3\,203\,226$ sites. The dotted lines in the insets correspond to the analytically predicted values $D_2 = 0.443$ and $\gamma = 0.416$. }
  \label{fig:partratio}
\end{figure}

In Figure \ref{fig:partratio} we show numerical results for the scaling exponent $\gamma$ of the average participation ratio according to Eq.\ \eref{equ:participation.1} for the Fibonacci chain and the labyrinth tilings. The data indicate that the scaling exponents for the labyrinth tiling in two and three dimensions approach the one-dimensional results for large system sizes \cite{JPhysCS.2010.Thiem, PhysRevB.2005.Cerovski}. We recently showed analytically that this property is fulfilled in the limit of infinite systems \cite{EPJB.2011.Thiem}. Additionally, for all systems the scaling exponents $\gamma$ approach a constant value for $w \to 0$. This limit behaviour can be related to the hierarchical structure of the chains and is further discussed in Sec.\ \ref{subsec:limitDq}.

\begin{figure}[b!]
 \centering
 \includegraphics[width=8cm]{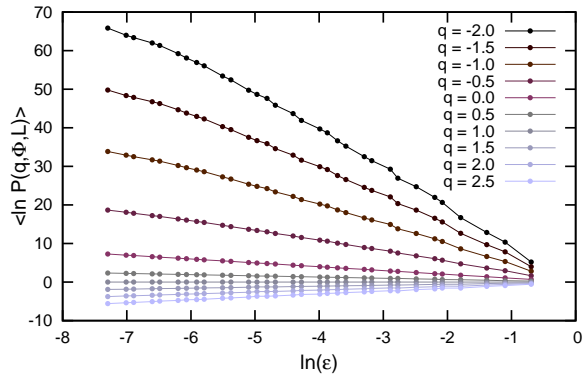}
 \caption{Scaling behaviour of the function $\langle \ln P(q,\Phi,L) \rangle$ used in the multifractal analysis with the ratio $\varepsilon$ for the Fibonacci chain $\mathcal{C}_{21}$ for $w=0.1$ and $s=1$.}
 \label{fig:multanalysis}
\end{figure}

\begin{figure}[t!]
 \centering
 \includegraphics[width=8cm]{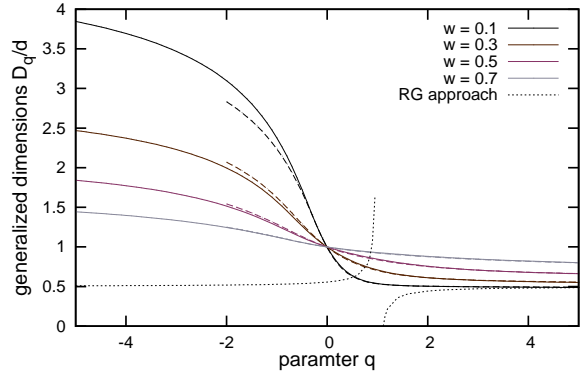}
 \caption{Scaling exponent $D_q/d$ determined via a multifractal analysis for the Fibonacci chain $\mathcal{C}_{21}$ (solid lines) and the two-dimensional labyrinth tiling $\mathcal{L}_{15}^{2{\rm d}}$ (dashed lines) for $s=1$.}
 \label{fig:DqPhi}
\end{figure}

The multifractal properties of the wave functions are obtained by a multifractal analysis described in Sec.\ \ref{subsec:def-part-ratio}. The scaling behaviour of the quantity $\langle \ln P(q,\Phi,L) \rangle$ with the box size $L$ is visualized for the Fibonacci chain in Figure \ref{fig:multanalysis}. For most of the systems the results can be well described by a power law. Of course, for large negative values of $q$ the accuracy deteriorates. A similar scaling behaviour can be observed for other values of $w$ and for the higher-dimensional systems although we find some deviations from the power law for higher-dimensional labyrinth tilings and small coupling parameters $w$. The corresponding results for the generalized dimensions $D_q$ are shown in Figure \ref{fig:DqPhi}. The results yield a box counting dimension of $D_0 = d$, which corresponds to the spatial dimension $d$ as expected. With increasing coupling parameter $w$ the range of the scaling exponents $D_q$ narrows, i.e., the strength of the multifractality is stronger for smaller coupling parameters $w$. While in the limit $w = s$ there is a single scaling exponent $D_q = d$, we find that for $w \to 0$ the distribution approaches a limit function, which can be related to the quasiperiodic structure. This is considered in more detail in the following section.

Further, the results in  Figure \ref{fig:DqPhi} indicate that the scaling exponents of different dimensions are related by $D_q^{d{\rm d}} = dD_q^{\rm 1d}$. This is not surprising as we have shown analytically that for the golden-mean labyrinth this relation is approached with increasing system size \cite{EPJB.2011.Thiem}.

%------------------------------------------------------------------------------------------------------------------------------------------
\subsection{Limit Behaviour of the Scaling Exponent $\gamma$ and the Generalized Dimensions $D_q$ for $w \to 0$}
\label{subsec:limitDq}
%------------------------------------------------------------------------------------------------------------------------------------------

If the coupling strength $w$ of the weak bond approaches zero, one can observe that the scaling exponent $\gamma$ of the participation ratio approaches a constant till it drops to zero for $w = 0$. This behaviour was already reported before \cite{PhysRevB.2005.Cerovski} but no explanation for this phenomena was given. A similar behaviour can also be found for the generalized dimensions $D_q$. This indicates that the wave functions do not become localized even for very small couplings $w$ and that they remain multifractals for $w \to 0$ regardless of the dimension. Thereby, this limit behaviour can be related to the hierarchical structure of the eigenstates of the quasiperiodic chains. In particular, for the Fibonacci chain $\mathcal{C}_a$ we can use the recursive structure of the energy spectrum according to Eq.\ \eref{equ:enegry-spectrum-fib} and of the wave functions according to Eqs.\ \eref{equ:wave-functions-atomRG} and \eref{equ:wave-functions-molRG} to derive the limit analytically.

We start by deriving the participation ratios for an atomic state ($f_{a-2} < i + f_{a-2} \le f_{a-1}$) of the Fibonacci chain
\numparts
\begin{eqnarray}
    p_a^{\rm atom} (\Psi^{i+f_{a-2}}, f_a)
     & = & \frac{1}{f_a} \left[ \sum_{l=1}^{f_a} \abs{\Psi_l^{i+f_{a-2}}}^4 \right]^{-1} \\
    &\stackrel{{\rm RG}}{\simeq} & \frac{1}{f_a} \left[ \tau^{-3} \sum_{l^\prime \in {\rm atom}} \abs{\Psi_{l^\prime}^i}^4 \right]^{-1}\\
     & = &\frac{f_{a-3}}{f_a} \tau^3 p_{a-3} (\Psi^i, f_{a-3})
      \simeq  p_{a-3} (\Psi^i, f_{a-3}) \;.
\end{eqnarray}
\endnumparts

Analogously, we obtain for the molecular states of the left main band of the energy spectrum ($1 \le i \le f_{a-2}$) the expression
\numparts
\begin{eqnarray}
    p_a^{\rm mol} (\Psi^i, f_a)
     & = & \frac{1}{f_a} \left[ \sum_{l=1}^{f_a} \abs{\Psi_l^i}^4 \right]^{-1} \\
     &\stackrel{{\rm RG}}{\simeq} &\frac{1}{f_a} \left[ \tau^{-2} \sum_{l^\prime \in {\rm mol}} \abs{\Psi_{l^\prime}^i}^4 \right]^{-1}\\
     &= & \frac{f_{a-2}}{f_a} \tau^2 \frac{1}{f_{a-2}} \left[ 2 \sum_{l^\prime=1}^{f_{a-2}} \abs{\Psi_{l^\prime}^i}^4 \right]^{-1} \\
     &\stackrel{f_a \gg 1}{\simeq} & \frac{1}{2} p_{a-2} (\Psi^i, f_{a-2}) \;.
\end{eqnarray}
\endnumparts

As these transformations do not directly depend on the energy we obtain the same expression also for the right main band of the energy spectrum. The average participation ratio is then directly calculated from these results according to
\numparts
\begin{eqnarray}
 \label{equ:rec-part-ratio}
    p_a(f_a)
   & = &  \frac{1}{f_a} \sum_{i=1}^{f_{a-2}} p_a^{\rm mol} (\Psi^i, f_a) + \frac{1}{f_a} \sum_{i=f_{a-2}+1}^{f_{a-1}} p_a^{\rm atom} (\Psi^i, f_a) \nonumber\\
   &~& \quad + \frac{1}{f_a} \sum_{i=f_{a-1}+1}^{f_{a}}  p_a^{\rm mol} (\Psi^i, f_a) \\
   & \stackrel{{\rm RG}}{\simeq} & \frac{f_{a-3}}{f_a} \frac{1}{f_{a-3}} \sum_{i=1}^{f_{a-3}}  p_{a-3} (\Psi^i, f_{a-3}) \nonumber\\
   &~& \quad + \frac{f_{a-2}}{f_a} \frac{1}{f_{a-2}} \sum_{i=1}^{f_{a-2}}  p_{a-2} (\Psi^i, f_{a-2}) \\
   & =  & \tau^{-3} p_{a-3} (f_{a-3})  + \tau^{-2} p_{a-2} (f_{a-2}) \;.
\end{eqnarray}
\endnumparts
With the scaling behaviour of the average participation ratio $p(f_a) \propto f_a^{-\gamma}$ according to Eq.\ \eref{equ:participation.1}, we obtain the relation
\begin{equation}
    1 = \tau^{-3(1+\gamma)} + \tau^{-2(1+\gamma)} \;.
\end{equation}
This expression is independent of the coupling parameter $w$ and yields a scaling exponent $\gamma \simeq 0.416$, which corresponds very well to the numerical results for $w \to 0$ in Figure \ref{fig:partratio}.

With the same approach we can also derive an analytical expression for the generalized dimensions $D_q$. Here, we start with the generalized inverse participation numbers for an atomic state according to Eq.\ \eref{equ:gIPN}, i.e.,
\numparts
\begin{eqnarray}
    Z_{q,a}^{\rm atom} (\Psi^{i+f_{a-2}})
     &= & \sum_{l=1}^{f_a} |\Psi_l^{i+f_{a-2}}|^{2q}  \\
     &\stackrel{{\rm RG}}{\simeq}& \tau^{-3q/2} \sum_{l^\prime \in {\rm atom}} |\Psi_{l^\prime}^i|^{2q}\\
     &= & \tau^{-3q/2} Z_{q,a-3} (\Psi^i)  \;.
\end{eqnarray}
\endnumparts
Likewise, we obtain for the molecular states of the left main band of the energy spectrum
\numparts
\begin{eqnarray}
    Z_{q,a}^{\rm mol}  (\Psi^i)
     &=& \sum_{l=1}^{f_a} |\Psi_l^i|^{2q}
     \stackrel{{\rm RG}}{\simeq}  \tau^{-q} \sum_{l^\prime \in {\rm mol}} |\Psi_{l^\prime}^i|^{2q} \\
     &=& 2 \tau^{-q} \sum_{l^\prime=1}^{f_{a-2}} |\Psi_{l^\prime}^i|^{2q} = 2 \tau^{-q}  Z_{q,a-2} (\Psi^i) \;.
\end{eqnarray}
\endnumparts
The same expression is again valid for the right main band of the energy spectrum. This yields the recursive equation for the average generalized inverse participation numbers
\numparts
\begin{eqnarray}
 \label{equ:rec-part-number}
    Z_{q,a}
   & = & \frac{1}{f_a} \sum_{i=1}^{f_{a-2}}  Z_{q,a}^{\rm mol} (\Psi^i) + \frac{1}{f_a} \sum_{i=f_{a-2}+1}^{f_{a-1}} Z_{q,a}^{\rm atom} (\Psi^i)  \nonumber \\
   & & \qquad + \frac{1}{f_a} \sum_{i=f_{a-1}+1}^{f_{a}}  Z_{q,a}^{\rm mol} (\Psi^i)\\
   & \stackrel{{\rm RG}}{\simeq} & \frac{f_{a-3}}{f_a} \frac{1}{f_{a-3}} \tau^{-3q/2} \sum_{i=1}^{f_{a-3}}  Z_{q,a-3} (\Psi^i)  \nonumber \\
   & & \qquad + 2 \frac{f_{a-2}}{f_a} \frac{1}{f_{a-2}} 2 \tau^{-q} \sum_{i=1}^{f_{a-2}}  Z_{q,a-2} (\Psi^i)\\
   & \stackrel{f_a \gg 1}{=} & \tau^{-3(q/2+1)} Z_{q,a-3}  + 4 \tau^{-(q+2)} Z_{q,a-2} \;.
\end{eqnarray}
\endnumparts
With the scaling behaviour of the average generalized inverse participation numbers $Z_q(f_a) \propto f_a^{-\tau_q}$ according to Eq.\ \eref{equ:gIPN}, we obtain the relation
\begin{equation}
 1 = \tau^{3(\tau_q - q/2 - 1)} + 4 \tau^{2(\tau_q - q/2 - 1)} \;.
\end{equation}
This expression is again independent of the coupling parameter $w$ and yields e.g.\ for $q=2$ the  scaling exponent $D_2 = \tau_2 \simeq 0.443$, which corresponds well to the numerical results for $w \to 0$ in Figure \ref{fig:partratio}. While this analytical result shows an even better agreement with the numerical results for larger values of $q$, there are significant differences for smaller values of $q$ as visualized in Figure \ref{fig:DqPhi}. This is caused by the fact that the measures for negative parameters $q$ are dominated by the boxes with the smallest probabilities. However, these probabilities are not considered in the RG approach because we neglect the probability on the other type of cluster. Nevertheless, the RG approach allows us to relate the quasiperiodic structure to the multifractal properties of the system for $q \ge 2$.

%******************************************************************************************************************************************
\section{Return Probability}
\label{sec:returnProb}
%******************************************************************************************************************************************

In order to obtain a deeper understanding of the connections between the transport properties and the quasiperiodic structure of a system, we study in this section the wave-packet dynamics for the different tilings. In the limit of weak coupling ($w \ll s$) it is again possible to connect the structure of the systems  and the wave-packet dynamics by an RG approach.

\mathversion{bold}
%------------------------------------------------------------------------------------------------------------------------------------------
\subsection{Scaling Behaviour of the Temporal Autocorrelation Function}
\label{subsec:autocorr}
%------------------------------------------------------------------------------------------------------------------------------------------
\mathversion{normal}

\begin{figure*}[t!]
 \centering
 \includegraphics[width=0.325\textwidth]{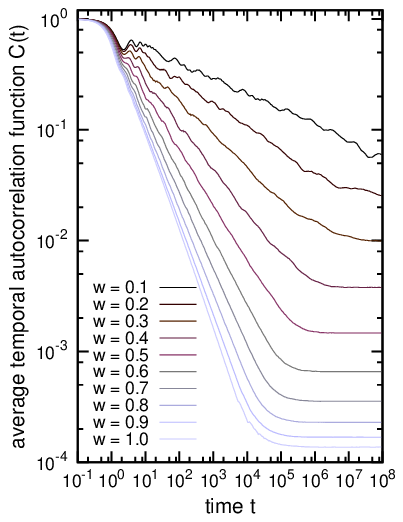}
 \includegraphics[width=0.325\textwidth]{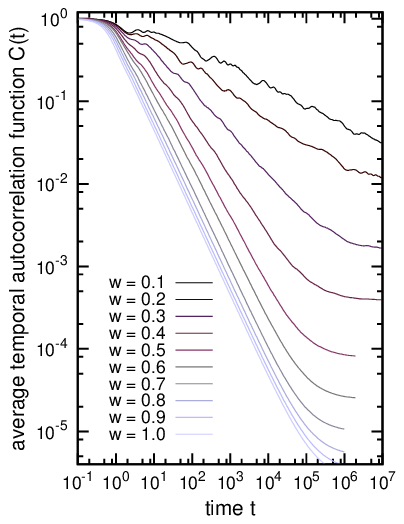}
 \includegraphics[width=0.325\textwidth]{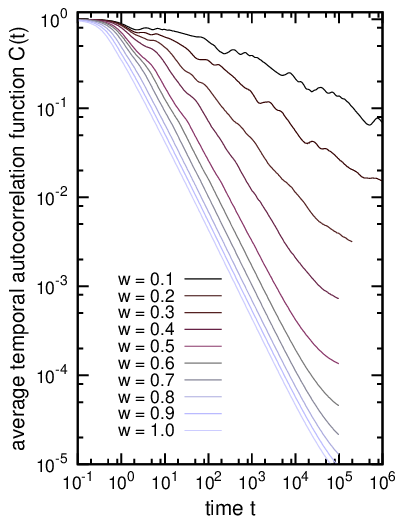}
 \caption{Average temporal autocorrelation function $C(t)$ for the Fibonacci chain $\mathcal{C}_{20}$ (left) and the associated labyrinth tilings $\mathcal{L}_{16}^{2{\rm d}}$ (center) and $\mathcal{L}_{12}^{\rm 3d}$ (right) in two and three dimensions for $s=1$.}
 \label{fig:autocorr}
\end{figure*}

\begin{figure*}[t!]
  \centering
  \includegraphics[width=0.49\textwidth]{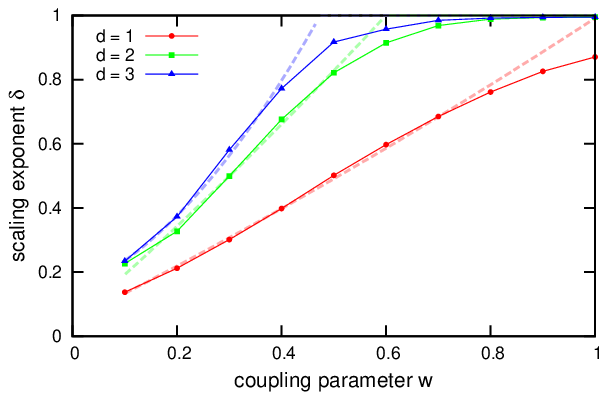}
  \includegraphics[width=0.49\textwidth]{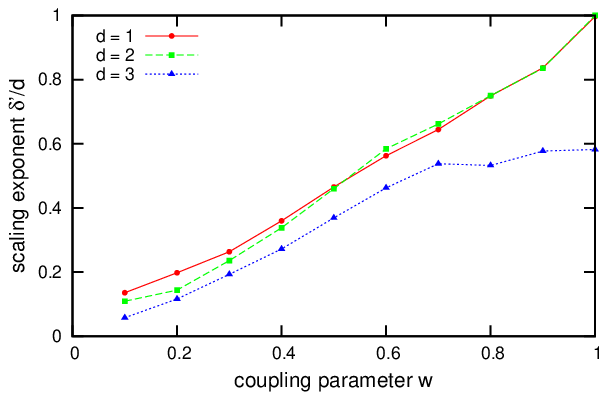}
  \caption{Scaling exponents $\delta$ of the temporal autocorrelation function and the expected limit behaviour (dashed lines) for the infinite systems (top) as well as the scaling exponent $\delta^\prime/d$ of the return probability $P(t)$ (bottom) averaged over different initial positions of the wave packet for the Fibonacci chain and the labyrinth tilings in two and three dimensions.}
  \label{fig:delta}
\end{figure*}

Some typical results for the temporal autocorrelation function $C(t)$ introduced in Sec.\ \ref{subsec:def-return-prob} are shown in Figure \ref{fig:autocorr} for the Fibonacci chain and the associated labyrinth tilings in two and three dimensions. The corresponding scaling exponents $\delta$ obtained by least squares fits are compiled in Figure \ref{fig:delta}. The power-law behaviour can be observed over several orders of magnitude in time, before $C(t)$ approaches a constant due to finite size effects. While we obtain $C(t) = 1$ for a system with localized wave functions, we expect to get $C(t \to \infty) \approx 1/N_a$ for a periodic chain of finite size $N_a$ due to the completely extended wave functions. A comparison with our numerical data for $w=s=1$ in Figure \ref{fig:autocorr} shows a good agreement with the latter result.

Further, the quantity $C(t)$ decreases more strongly for higher values of the coupling parameter $w$ in agreement with the more extended wave functions in these systems. However, for a periodic chain in one dimension one does not obtain the expected result $\delta = 1$ by a least squares fit due to sublogarithmic contributions \cite{JPhys.1995.Zhong}. Additionally, for small values of $w$ we observe some oscillations. This corresponds to a step-like behaviour of $C(\mathbf{r}_0,t)$, which can be related to the hierarchical structure of the chains \cite{PhysRevB.2009.Thiem}.

Comparing the scaling behaviour of the average temporal autocorrelation function $C(t)$ in Figure \ref{fig:autocorr} with respect to the dimension, we observe that for the one-dimensional chain the absolute value of the slope steadily increases with increasing coupling parameter $w$, while in two and three dimensions $C(t)$ is nearly identical for $w^{\rm 2d} > 0.6 s$ and $w^{\rm 3d} > 0.5 s$. These differences are also clearly visible for the corresponding scaling exponents $\delta$ shown in Figure \ref{fig:delta}, which approach one when the coupling parameter $w$ becomes larger than the threshold parameters $w_{\rm th}^{\rm 2d} \approx 0.6 s$ and $w_{\rm th}^{\rm 3d} \approx 0.45 s$ for which the spectrum becomes absolutely continuous. It is expected that $\delta = 1$ for infinite systems if $w > w_{\rm th}^{\rm 2d}$. This limit behaviour of the scaling exponent $\delta$ is indicated by the dashed lines in Figure \ref{fig:delta}. Similar results can be found for other metallic-mean systems \cite{PhysRevB.2005.Cerovski}. Although $\delta \to 1$ indicates ballistic transport for the one-dimensional periodic chains \cite{JPhys.1995.Zhong}, the behaviour in two and three dimensions does not correspond to a transition towards ballistic spreading. The latter result is not always correctly used in literature because e.g.\ Zhong and Mosseri wrongly state that there is a crossover from ballistic-like behaviour to a non-ballistic-like behaviour in the hypercubic tiling \cite{JPhys.1995.Zhong}. The results for the mean square displacement of the wave packet for the hypercubic or the labyrinth tiling clearly reveal anomalous transport for $w < s$. Only for $w=s$ ballistic transport is observed \cite{PhysRevB.2009.Thiem, PhysRevB.2000.Yuan}.

\mathversion{bold}
%------------------------------------------------------------------------------------------------------------------------------------------
\subsection{Scaling Behaviour of the Return Probability}
\label{subsec:return-prob}
%------------------------------------------------------------------------------------------------------------------------------------------
\mathversion{normal}

\begin{figure}[b!]
  \centering
  \includegraphics[width=8cm]{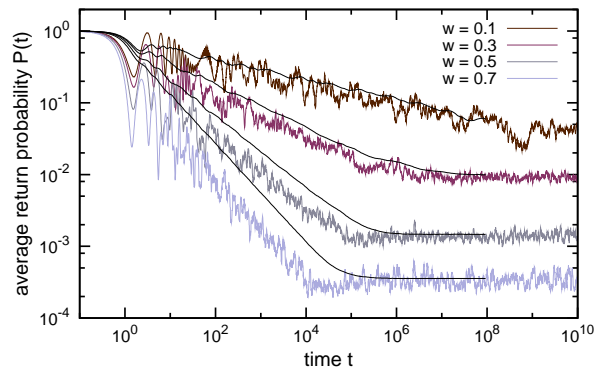}
  \caption{Moving average of the return probability $P(t)$ over 10 data points (light colors) and the corresponding temporal autocorrelation function $C(t)$ (dark colors) for the Fibonacci chain $\mathcal{C}_{20}$. Both quantities are averaged over 500 initial positions of the wave packet.}
  \label{fig:paca-smooth}
\end{figure}

\begin{figure}[t!]
  \centering
  \includegraphics[width=8cm]{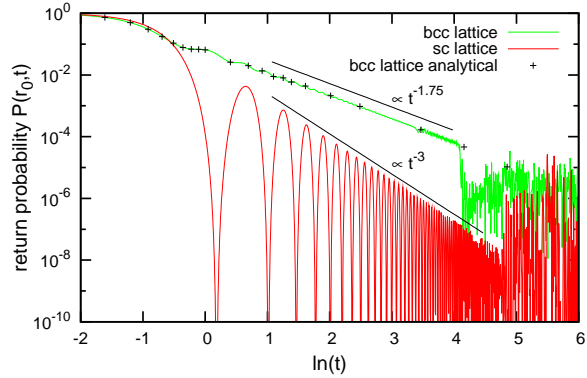}
  \caption{Scaling behaviour of the return probability $P(\mathbf{r}_0,t)$ for the bcc lattice with $240^3/4$ sites and the sc lattice with $240^3$ sites as well as the analytical results of the return probability $P(\mathbf{r}_0,t)$ for the bcc lattice according to Eq.\ \eref{equ:paca-bcc}.}
  \label{fig:paca-cmp}
\end{figure}

In order to obtain information about transport properties instead of the properties of the energy spectrum we can make use of the fact that already the integrand $P(t)$ in Eq.\ \eref{equ:autocorrelation} shows a power-law behaviour according to $P(t) \propto t^{-\delta^\prime}$. The integration itself is only used for smoothing the results because $P(t)$ shows considerable oscillations (cf.\ e.g.\ Figs.\ \ref{fig:paca-smooth} and \ref{fig:paca-cmp}). Naturally, one would expect that the scaling exponents of $C(t)$ and $P(t)$ are equal, i.e., $\delta = \delta^\prime$. However, the integration leads to two basic disadvantages. For $\delta^\prime = 1$ we obtain an additional logarithmic contribution $C(t) \propto \ln(t) / t$ from the integration \cite{JPhys.1995.Zhong}. Hence, for the periodic one-dimensional systems and the commonly used system sizes one obtains exponents $\delta$ significantly lower than $1$ in agreement with our result $\delta \approx 0.87$. Further, in higher dimensions the scaling exponent $\delta^\prime$ of $P(t)$ becomes greater than 1 for large coupling parameters $w$ (cf.\ Figure \ref{fig:delta}). Zhong and Mosseri pointed out that in this case the integral in Eq.\ \eref{equ:autocorrelation} converges and yields a constant factor, which results in $C(t) \propto 1/t$. Hence, the scaling exponent $\delta = 1$ is obtained in the absolute continuous regime \cite{JPhys.1995.Zhong}. From Figure \ref{fig:delta} we find $\delta^\prime > 1$ for $w > 0.55 s$ in two dimensions and $w > 0.43 s$ in three dimensions in reasonable agreement with $w_{\rm th}^{\rm 2d}$ and $w_{\rm th}^{\rm 3d}$. Since the integral converges rather slowly for coupling parameters $w$ only slightly larger than $w_{\rm th}$, we do not observe $\delta = 1$ for the finite systems considered here. However, for an infinite system we expect to obtain a clear transition towards $\delta = 1$ as indicated by the dashed lines in Figure \ref{fig:delta}.

Therefore, we lose information about the transport properties and $\delta = 1$ is misleading because it often does not occur where it is expected (e.g.\ for periodic one-dimensional systems) but wrongly suggests a ballistic behaviour for $w > w_{\rm th}$ in two and three dimensions. Hence, for studying the transport properties it is better to extract the scaling exponent $\delta^\prime$ via the calculation of $P(t)$, or the temporal autocorrelation has to be redefined so that the integrand does not converge anymore. We choose the first option and use a moving average instead of the integration to smooth the results as shown in Figure \ref{fig:paca-smooth}.

In Figure \ref{fig:delta} the scaling exponent $\delta^\prime$ of the return probability $P(t)$ divided by the spatial dimension $d$ is shown for the Fibonacci chain and the labyrinth tilings. The comparison of the scaling exponents for the temporal autocorrelation $C(t)$ and its integrand $P(t)$ shows that both results are quite close for the one-dimensional systems. The main difference occurs for $w > 0.8 s$, where we obtain the expected value $\delta^\prime \approx 1$ for $w=s$ while the evaluation of $C(t)$ results in $\delta \approx 0.87$. Further, there are some differences for the Fibonacci chain, especially, for small values of $w$. This probably originates from the fact that we have not considered large enough times in the calculations of $C(t)$. Actually, it is much more time-consuming to compute $C(t)$ because the integration requires the consideration of much smaller time intervals. This is another advantage of calculating only the quantity $P(t)$.

Further, in two dimensions we observe that the scaling exponents $\delta_{\rm 2d}^\prime$ are about twice as large as the one-dimensional scaling exponents $\delta_{\rm 1d}^\prime$. This is in good agreement with the results for the hypercubic tiling in Eq.\ \eref{equ:deltaprime-nD}. However, the results for the three-dimensional labyrinth tiling do not entirely fit into this scheme. While for $w \le 0.6 s$ the expression $\delta_{\rm 3d}^\prime/d$ is only slightly smaller than the exponent $\delta_{\rm 1d}^\prime$, the behaviour changes completely for large values of $w$. In the latter case the scaling exponent $\delta_{\rm 3d}^\prime$ becomes almost constant and approaches $\delta_{\rm 3d}^\prime(w=s) \approx 1.75$. The exact reason for this behaviour is unclear yet. Having a look at $P(\mathbf{r}_0,t)$ for different initial positions of the wave packet for $w=s$, one can observe that after an unusually smooth decrease of the function $P(\mathbf{r}_0,t)$ according to a power law, there is an abrupt dip in the curves when finite size effects first occur (cf.\ Figure \ref{fig:paca-cmp}).

\begin{figure}[t!]
  \centering
  \includegraphics[width=4cm]{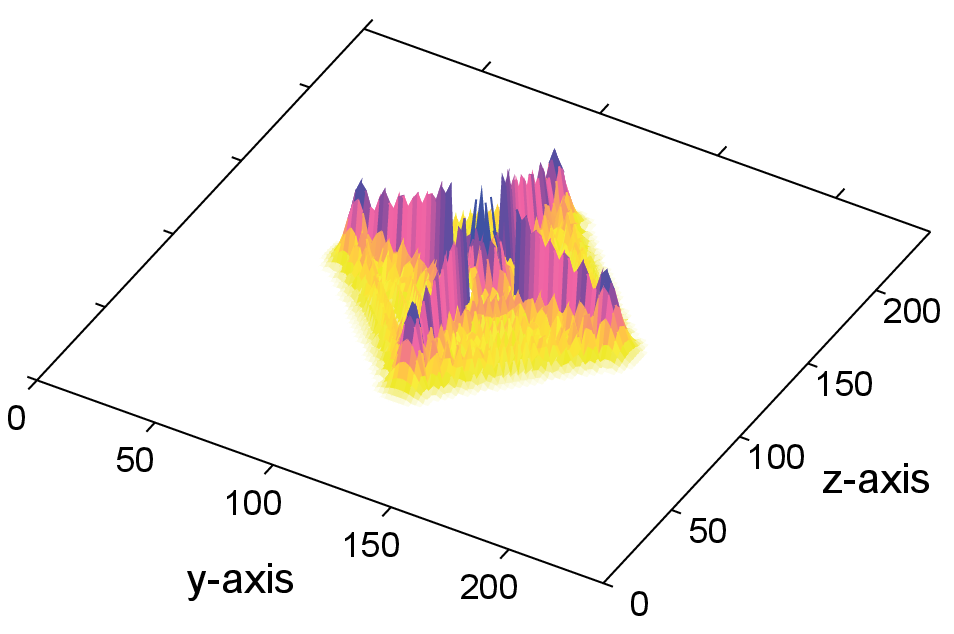}
  \includegraphics[width=4cm]{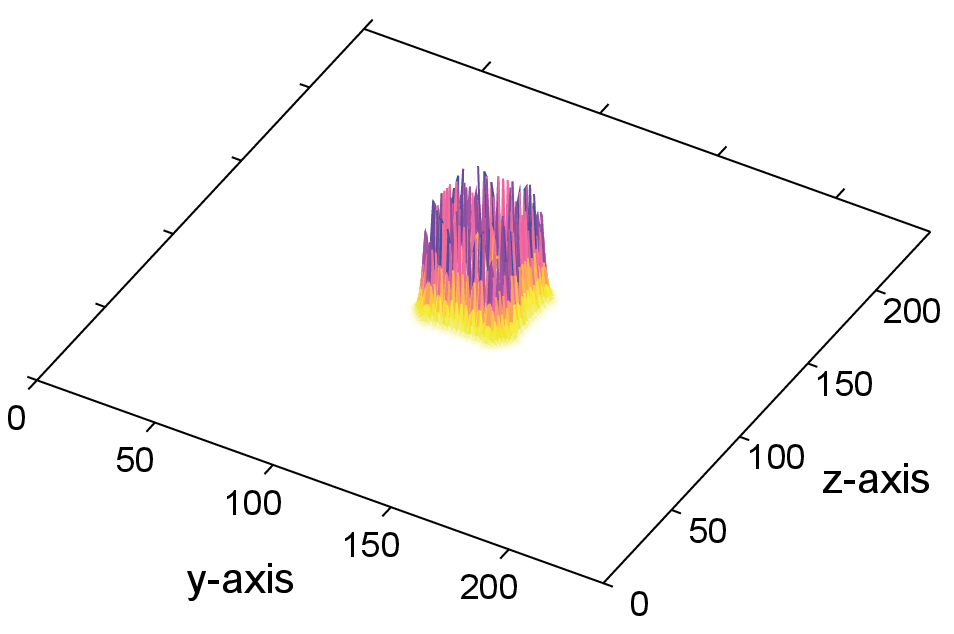}\\
  \includegraphics[width=4cm]{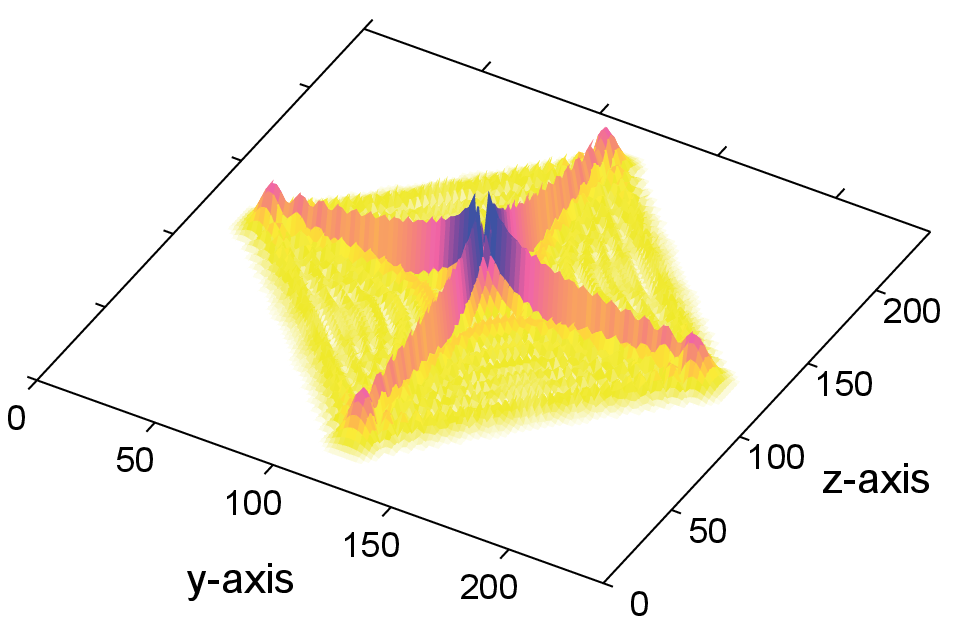}
  \includegraphics[width=4cm]{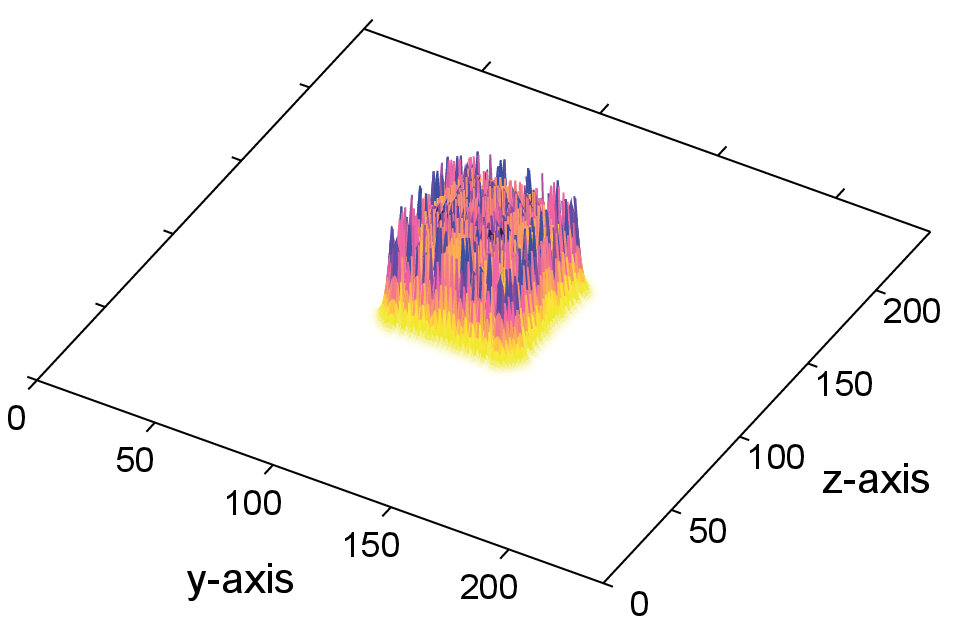}\\
  \includegraphics[width=4cm]{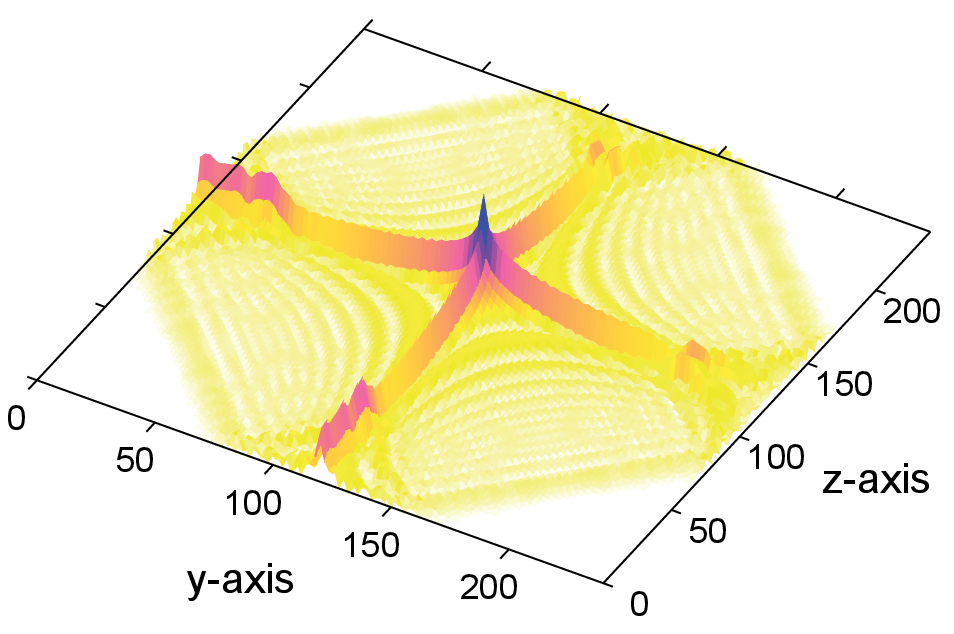}
  \includegraphics[width=4cm]{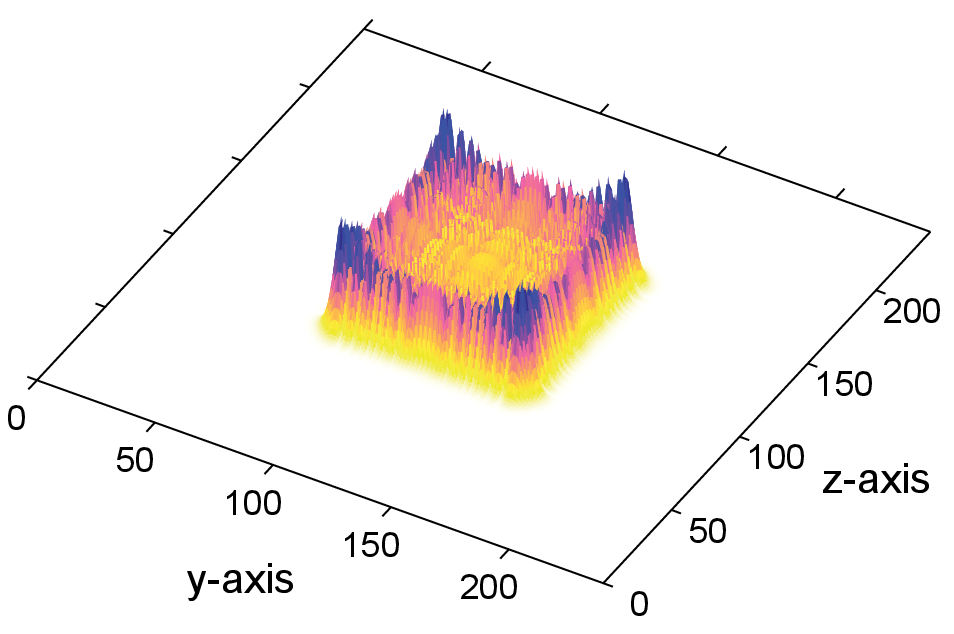}
  \caption{Probability density along a plane through the center of the bcc lattice with $240^3/4$ sites (left column) and the sc lattice with $240^3$ sites (right column) at the times $\ln t = 2$ (top), $\ln t = 2.5$ (center), and $\ln t = 3$ (bottom). The corresponding values of $P(\mathbf{r}_0,t)$ are shown in Figure \ref{fig:paca-cmp}.}
  \label{fig:paca3D2}
\end{figure}

Since the square Fibonacci tiling and the two-dimensional labyrinth tiling are also identical for $w=s$, the scaling exponents $\delta$ and $\delta^\prime$ are identical. However, in three dimensions for $w \to s$ the labyrinth tiling becomes a \emph{body centered cubic} (bcc) lattice while the cubic tiling approaches a \emph{simple cubic} (sc) lattice. Therefore, these two grids have a different number of nearest neighbours independent of the coupling parameter $w$, i.e., eight in the labyrinth tiling and six in the cubic tiling.
While the numerical results for the two-dimensional labyrinth tiling satisfy the relations in Eqs.\ \eref{equ:deltaprime-nD} and \eref{equ:delta-nD}, there are significant differences in the wave-packet dynamics in three dimensions, which are probably related to the different grid structure. Therefore, in Figure \ref{fig:paca3D2} we plot for both systems the probability density of a three-dimensional wave packet along the $y$-$z$ plane through the center of the $x$ axis for different times $t$ and the corresponding return probability $P(\mathbf{r}_0,t)$ in Figure \ref{fig:paca-cmp}. While the probability density of the wave packet for the sc lattice is distributed in a cube around the origin, for the bcc lattice it spreads in the shape of an octahedron from the center with some prominent resonances along the $x$, $y$, and $z$ directions. For the labyrinth tiling the wave packet already reached the boundaries for $\ln t = 3$ and was reflected due to the free boundary conditions. This leads to constructive interference at two boundaries and to destructive interference at the opposite boundaries because the initial position of the wave packet is chosen from one of the four possible center positions of the labyrinth for even chain lengths, which results in a slight asymmetry of the wave-packet dynamics.

According to Figure \ref{fig:paca-cmp} the return probability for the sc lattice strongly oscillates and decreases with a scaling exponent of $\delta^\prime = 3$, while for the bcc lattice it decreases rather smoothly with a scaling exponent $\delta^\prime \approx 1.75$. The dip for $\ln{t} > 4$ for the bcc structure is clearly due to finite size effects and thus has not to be considered for the fitting of the scaling exponent. Although the center of the wave packet decreases more slowly for the bcc structure than for the sc lattice, the overall wave packet spreads significantly faster for the bcc lattice. This might be related to the higher number of nearest neighbours in the bcc lattice.

The data for the sc lattice are in good agreement with the results of Zhong and Mosseri (cf.\ Sec.\ \ref{subsec:def-return-prob}). Their approach can also be used for the determination of the return probability $P(t) = \left|\Upsilon_{\mathbf{r}_0} (t) \right|^2 $ for the bcc lattice. The approach makes use of the energy spectrum of the bcc lattice in $\mathbf{k}$ space \cite{JPhysChem.1969.Jellito}
\begin{equation}
    E(\mathbf{k}) = 8 \cos{\frac{k_x}{2}} \cos{\frac{k_y}{2}} \cos{\frac{k_z}{2}} \;.
\end{equation}
The probability amplitude at the initial position $\mathbf{r}_0$ of the wave packet can be determined by the Fourier transform of this energy spectrum. Hence, we obtain
\numparts
\begin{eqnarray}
\label{equ:paca-bcc}
    & &\Upsilon_{\mathbf{r}_0} (\mathbf{r}_0, t) = \int \rme^{-\rmi E t} \rmd \mu(E) = \int_{\rm BZ} e^{-\rmi E(\mathbf{k}) t} \rmd \mathbf{k} \\
                            &=& \frac{1}{(2\pi)^3} \int\limits_0^{2\pi} \int\limits_0^{2\pi} \int\limits_0^{2\pi}  \rme ^ {\left( - \rmi 8 t \cos{\frac{k_x}{2}} \cos{\frac{k_y}{2}} \cos{\frac{k_z}{2}} \right)}  \rmd k_x  \rmd k_y \rmd k_z \\
                            &=& \frac{1}{(2\pi)^2} \int\limits_0^{2\pi} \int\limits_0^{2\pi} J_0 \left( 8 t \cos{\frac{k_y}{2}} \cos{\frac{k_z}{2}} \right) \rmd k_y \rmd k_z \;.
\end{eqnarray}
\endnumparts

We use the fact that the integration over the Brillouin zone (BZ) can be reduced to an integral over a cube from $0$ to $2\pi$ due to the symmetries of the components of the $\mathbf{k}$ vector \cite{JPhysChem.1969.Jellito}. The function $J_0$ denotes the Bessel function of the first kind.
The integral over the Bessel function is evaluated for various time steps by numerical integration. The results agree exactly with the numerical data as shown in Figure \ref{fig:paca-cmp}. Unfortunately, it is not possible to evaluate this integral for arbitrary large times because the Bessel function strongly oscillates, which leads to numerical problems. While it is possible to derive with this approach an analytical expression for the long-time behaviour of the return probability for the sc lattice, we were not able to obtain a corresponding result for the bcc lattice.

%------------------------------------------------------------------------------------------------------------------------------------------
\subsection{RG Approach for the Fibonacci Chain}
\label{subsec:RG-theory-return-prob}
%------------------------------------------------------------------------------------------------------------------------------------------

While we considered the limit case of the periodic systems in the previous section, we can also find a connection between the structure of the Fibonacci chain $\mathcal{C}$ and the transport properties in the regime of strong quasiperiodic modulation. In particular, we are able to derive an analytical expression for the scaling exponents $\delta^\prime$ of the return probability $P(t)$. This is based on the results of the RG approach, which yields a scaling behaviour of the energy spectrum according to Eq.\ \eref{equ:enegry-spectrum-fib} and of the wave functions according to Eqs.\ \eref{equ:wave-functions-atomRG} and \eref{equ:wave-functions-molRG}.

First, we determine the return probability $P_a(l_0,t)$ for an approximant $a$ for the two different kinds of initial positions $l_0$. For $l_0$ being an atomic site we obtain
\numparts
\begin{eqnarray}
 P_a^{\rm atom}(l_0,t)
    & \stackrel{\eref{equ:return-prob}}{=}& \left|\Upsilon_{l_0}^{\rm atom} (l_0,t) \right|^2 \\
    & \stackrel{\eref{equ:wave-pacekt}}{\simeq}& \left| \sum_{i=f_{a-2}+1}^{f_{a-1}} e^{-\rmi E_a^i t } \left| \Psi_{l_0,a}^i \right|^2 \right|^2 \\
    & \stackrel{\rm RG}{\simeq}& \left| \sum_{i=1}^{f_{a-3}} e^{-\rmi E_{a-3}^i \bar{z} t }  {\tau^{-\frac{3}{2}}}  \left| \Psi_{l_0^\prime, a-3 }^i \right|^2 \right|^2 \\
    & = & \tau^{-3} P_{a-3}(l_0^\prime,\bar{z} t)\;,
\end{eqnarray}
\endnumparts
and for $l_0$ belonging to a molecular site we get
\numparts
\begin{eqnarray}
 P_a^{\rm mol}(l_0,t)
    &\stackrel{\eref{equ:wave-pacekt}}{\simeq} & \left| \sum_{i=1}^{f_{a-2}} e^{-\rmi E_a^i t } \left|\Psi_{l_0,a}^i \right|^2 + \sum_{i=f_{a-1}+1}^{f_a} e^{-\rmi E_a^i t } \left| \Psi_{l_0,a}^i  \right|^2 \right|^2 \\
    &\stackrel{\rm RG}{\simeq} & \left| \sum_{i=1}^{f_{a-2}} e^{-\rmi E_{a-2}^i z t } \tau^{-1}  \left|\Psi_{l_0^\prime, a-2}^i \right|^2 \left (  e^{ \rmi s t } +  e^{ -\rmi s t } \right) \right|^2 \\
    & = & 4 \cos^2 (st) \tau^{-2} P_{a-2}(l_0^\prime,z t) \\
    &\stackrel{t \to \infty}{\simeq}& 2 \tau^{-2} P_{a-2}(l_0^\prime,z t) \;.
\end{eqnarray}
\endnumparts
In the last step we make use of the fact that for large times $t$ only the average behaviour of $\cos^2 (st)$ is relevant, which results in a factor of $1/2$. The average return probability for an approximant $a$ is then given by
\numparts
\begin{eqnarray}
 P_a(t) & = & \frac{1}{f_a} \sum_{l_0=1}^{f_a} P_a(l_0,t) \\
     &= & \frac{1}{f_a} \sum_{l_0 \in {\rm atom}} P_a^{\rm atom}(l_0,t) + \frac{1}{f_a} \sum_{l_0 \in {\rm mol}} P_a^{\rm mol}(l_0,t) \\
     &\stackrel{\rm RG}{\simeq} & \frac{f_{a-3}}{f_a} \tau^{-3} \frac{1}{f_{a-3}}  \sum_{l_0^\prime=1}^{f_{a-3}} P_{a-3}(l_0^\prime,\bar{z} t) \nonumber\\
     & \, & \qquad+  2 \frac{f_{a-2}}{f_a} 2 \tau^{-2} \frac{1}{f_{a-2}} \sum_{l_0^\prime=1}^{f_{a-2}} P_{a-2}(l_0^\prime,z t) \\
     &\stackrel{f_a \gg 1}{\simeq}&  \tau^{-6} P_{a-3}(\bar{z} t) + 4 \tau^{-4} P_{a-2}(z t) \;.
\end{eqnarray}
\endnumparts

\begin{figure}[t!]
  \centering
  \includegraphics[width=8cm]{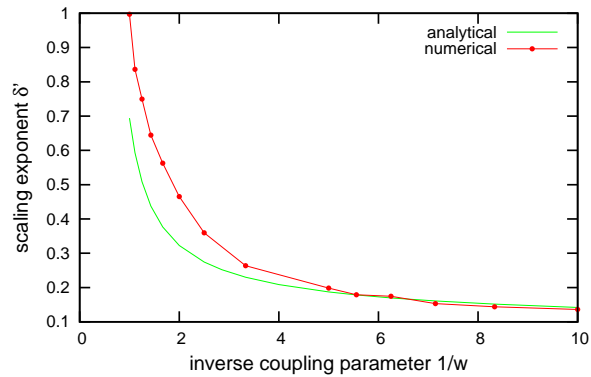}
  \caption{Comparison of the analytical and the numerical results of the scaling exponent $\delta^\prime$ for the Fibonacci chain $\mathcal{C}_{20}$.}
  \label{fig:paca-cmp2}
\end{figure}

With the scaling behaviour of the return probability $P(t) \propto t^{-\delta^\prime}$, we obtain the equation
\begin{equation}
    1 = \tau^{-6} \bar{z}^{-\delta^\prime} + 4 \tau^{-4} z^{-\delta^\prime}\;,
\end{equation}
which approaches the numerical results for decreasing values of $w$ as shown in Figure \ref{fig:paca-cmp2}. Note that Pi\'echon \cite{PhysRevLett.1996.Piechon} studied a similar quantity, which he also denoted as return probability. However, that quantity is related to the spectral measure of the density of states. In contrast, the return probability $P(t)$ is related to the spectral measure of the local density of states for the initial position $l_0$ of the wave packet. Only the latter quantity is related to the wave-packet dynamics.

%------------------------------------------------------------------------------------------------------------------------------------------
\section{Lower Bound for Scaling Exponent of the Mean Square Displacement}
\label{subsec:lower-bound}
%------------------------------------------------------------------------------------------------------------------------------------------

Another quantity often considered for the description of the electronic transport properties is the mean square displacement
(also called width)
\begin{equation}
 d(\mathbf{r}_0,t) = \left[\sum_{\mathbf{r} \in \mathcal{L}} |\mathbf{r}-\mathbf{r}_0|^2 \, |\Upsilon_\mathbf{r}(\mathbf{r}_0,t)|^2 \right]^{\frac{1}{2}}
\end{equation}
of the wave packet.
The spreading of the width $d(\mathbf{r}_0,t)$ of the wave packet for an infinite system shows anomalous diffusion for $t \rightarrow \infty$ according to $d(\mathbf{r}_0, t) \propto t^{\beta(\mathbf{r}_0)}$, where the scaling exponent $\beta(\mathbf{r}_0)$ depends on the initial position $\mathbf{r}_0$ of the wave packet \cite{RevMathPhys.1998.SchulzBaldes, JMathPhys.1997.Roche, AdvPhys.1992.Poon, PhysRevLett.1994.Huckestein, PhysRevB.2000.Yuan}. The electronic transport properties are governed by the wave-packet dynamics averaged over different initial positions, i.e., $d(t) = \langle d(\mathbf{r}_0,t) \rangle \propto t^{\beta}$. Thereby, the scaling exponent $\beta$ is related to the conductivity $\sigma$ via the generalized Drude formula for zero-frequency conductivity $\sigma \simeq  e^2 \varrho(E_{\rm F}) c t_{\rm sca}^{2\beta -1}$ \cite{JMathPhys.1997.Roche, PhysRevLett.1997.Schulz-Baldes}.
The quantity $e$ denotes the elementary charge of an electron, $\varrho(E_{\rm F})$ the density of states at the Fermi level $E_{\rm F}$, $c$ a constant, and $t_{\rm sca}$ a characteristic time beyond which propagation becomes diffusive due to scattering \cite{PhysRevLett.2000.Mayou, RevMathPhys.1998.SchulzBaldes}. Hence, $\beta=0$ corresponds to the absence of diffusion, $\beta=1/2$ to classical diffusion, and $\beta=1$ to ballistic spreading. For quasiperiodic structures one often observes anomalous diffusion characterized by $0 < \beta < 1 $ \cite{PhysRevB.1992.Passaro, JPhysFrance.1989.Sire, PhysRevB.2000.Yuan}. For $w \ll s$ the width of the wave packet can be related by an RG approach to the structure of the Fibonacci chain \cite{PhysRevLett.1996.Piechon} and the labyrinth tiling \cite{PhysRevB.2012.Thiem}.

The computation of the scaling exponent $\beta$ is computationally very expensive especially in higher dimensions because it includes an average over all sites of the system. However, for the hypercubic tiling the scaling exponents are related according to $ \beta_{d{\rm d}} = \beta_{\rm 1d} $ due to the separability of the time-evolution operator. Numerical and analytical results suggest that this relation could be also valid for the labyrinth tilings \cite{PhysRevB.2000.Yuan, PhysRevB.2005.Cerovski,PhysRevB.2012.Thiem}.
However, in general one can make use of several lower bounds for the scaling exponents $\beta$, which have been proposed in literature. We shortly introduce two of them here and compare them to our numerical data \cite{JMathAA.1997.Barbaroux, EurophysLett.1993.Guarneri, PhysRevLett.1997.Ketzmerick, PhysRevB.2000.Yuan}. Further, we present an improved lower bound for systems with absolute continuous energy spectra.

\begin{figure*}[t!]
  \centering
  \includegraphics[width=0.328\textwidth]{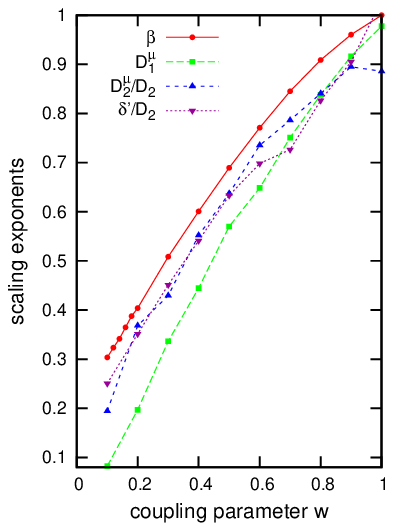}
  \includegraphics[width=0.328\textwidth]{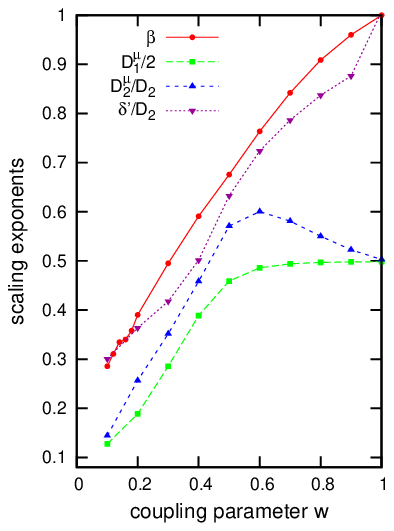}
  \includegraphics[width=0.328\textwidth]{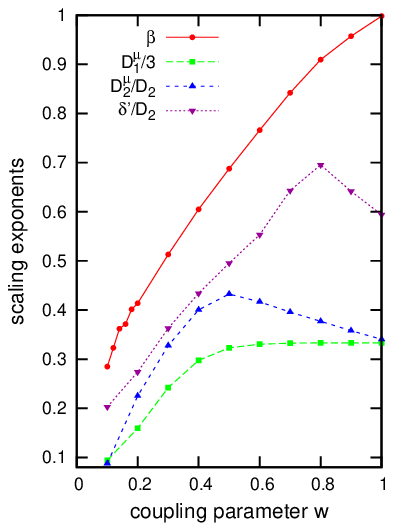}
  \caption{Comparison of the scaling exponent $\beta$ with the lower bounds $D_1^\mu / d$, $D_2^\mu / D_2$, and $\delta^\prime / D_2$ for the Fibonacci chain (left) and the corresponding labyrinth tilings in two dimensions (center) and three dimensions (right).}
  \label{fig:bounds-beta}
\end{figure*}

\emph{Lower bound by Guarneri:} As a rule of thumb, the wave-packet propagation is faster for smoother spectral measures. In particular, Guarneri proved for the scaling exponent $\beta$ that in one dimension the lower bound $\beta \ge D_1^{\mu}$ holds \cite{EurophysLett.1993.Guarneri}. This can be generalized for arbitrary dimensions to
\begin{equation}
 \label{equ:guarneri-inequality}
    \beta \ge \frac{D_1^\mu}{d} \;.
\end{equation}
This expression quantitatively relates the diffusive properties to the information dimension of the spectral measure of the local density of states. A comparison with our numerical results in Figure \ref{fig:bounds-beta} shows that this inequality is clearly satisfied in one dimension.
The spectral dimensions $D_q^\mu$ are computed by a multifractal analysis with respect to the energy. Therefore, we divide the energy range into $B$ boxes of size $\Delta E$ and compute for a certain vertex $\mathbf{r}_0$ of the tiling the function
\begin{equation}
 \label{equ:partfuncLDOS}
 \Gamma(\Delta E) = \left\langle \sum_{b=1}^{B} \left( \sum_{E^\mathbf{s} \in \,{\rm box}\, b} \abs{\Phi_{\mathbf{r}_0}^\mathbf{s}}^2 \right)^q \right\rangle \propto \left(\Delta E\right)^{D_q^\mu}\;,
\end{equation}
which shows a power-law behaviour with the generalized spectral dimension $D_q^\mu$ \cite{APhys.1996.Brandes, PhysRevA.1986.Halsey}.

With the inequality in Eq.\ \eref{equ:guarneri-inequality} it is possible to show that an absolutely continuous energy spectrum in one dimension implies ballistic transport \cite{JStatPhys.2000.Bellissard} and that in two dimensions this inequality requires a superdiffusive evolution of the wave packet ($\beta > \frac{1}{2}$) in the presence of an absolutely continuous spectrum. However, in three dimensions it allows the occurrence of subdiffusive wave-packet spreading ($\beta < \frac{1}{2}$) although the energy spectrum is purely absolutely continuous, i.e., $D_1^\mu = 1$ \cite{JStatPhys.2000.Bellissard}. This is a very interesting aspect because investigations on real quasicrystals revealed that they are usually accompanied by absolutely continuous spectra \cite{JStatPhys.2000.Bellissard} but the conductivity is well described by the generalized Drude formula for the case $\beta < \frac{1}{2}$ \cite{PhysRevLett.1993.Mayou, AdvPhys.1992.Poon, JMathPhys.1997.Roche, PhysRevLett.1997.Schulz-Baldes}. Since $D_q^{\mu,d{\rm d}}$ eventually becomes $1$ for an absolutely continuous spectrum, this bound strongly underestimates the actual scaling exponent $\beta$ in two and three dimensions for large values of $w$ as shown in Figure \ref{fig:bounds-beta}.

\emph{Lower bound by Ketzmerick et al.:} Using the fact that the center of the wave packet is known \cite{JPhysMath.1996.Guarneri} to decay according to $t^{-D_2^\mu}$, Ketzmerick et al.\ showed that due to the normalization condition the spreading of a wave packet is described by an exponent $\beta = D_2^\mu/d$ in $d$ dimensions as long as $D_2^\mu < 1$, i.e., for singular continuous spectra \cite{PhysRevLett.1997.Ketzmerick}. For wave packets which spread in a space of reduced dimension $D_2$ instead of $d$ dimensions they showed that the relation
\begin{equation}
 \label{equ:equ-ketzmerick}
    \beta = \frac{D_2^\mu}{D_2}
\end{equation}
holds for all systems which are characterized by a single scaling exponent $\beta = \beta_q$ \cite{PhysRevLett.1997.Ketzmerick}. For the more general case of multiscaling in time ($d_q(t) = \sum_{\mathbf{r} \ne \mathbf{r}_0} \abs{\mathbf{r}-\mathbf{r}_0}^q \, |\Upsilon_\mathbf{r}(\mathbf{r}_0,t)|^2 \propto t^{q\beta_q}$) they found that this yields a lower bound for positive moments, i.e.,
\begin{equation}
 \label{equ:inequ-ketzmerick}
    \beta_q \ge \frac{D_2^\mu}{D_2} \;.
\end{equation}
For negative moments one obtains a respective upper bound.

Comparing this with our numerical results for $q=2$ in Figure \ref{fig:bounds-beta} for the one-dimensional case, we find that the inequality in Eq.\ \eref{equ:inequ-ketzmerick} is fulfilled and provides a good lower bound, which is significantly better than the result by Guarneri in Eq.\ \eref{equ:guarneri-inequality} \cite{EurophysLett.1993.Guarneri}. The largest differences occur for $w \to s$, which are partly caused by numerical issues because the relation $D_2^\mu = 1$ is satisfied for a periodic chain. The numerical results only fulfill the lower bound according to Eq.\ \eref{equ:inequ-ketzmerick} because the wave-packet dynamics for the Fibonacci chain show multiscaling in time.
This inequality also holds in two and three dimensions.
However, in the regime of an absolutely continuous spectrum, the spectral dimensions fulfill $D_q^\mu = 1$, and the inequality in Eq.\ \eref{equ:inequ-ketzmerick} is no longer a good lower bound as shown in Figure \ref{fig:bounds-beta}.

Addressing the latter case we would like to point out that the decay of the center of the wave packet is only described by the correlation dimensions of the local density of states for $w < w_{\rm th}$ because for $w > w_{\rm th}$ we obtain $D_2^\mu = 1$ and the integral of the autocorrelation function  converges \cite{JPhys.1995.Zhong}. However, in this regime the scaling exponent $\delta^\prime$, which describes the decay of the center of the wave packet, can become larger than 1 (cf.\ Figure \ref{fig:delta}). Hence, for $d$-dimensional systems the bound in Eq.\ \eref{equ:inequ-ketzmerick} should be replaced by
\begin{equation}
  \label{equ:inequ-deltap}
    \beta \ge  \frac{\delta^\prime}{D_2}
\end{equation}
in the absolutely continuous regime. According to Eq.\ \eref{equ:deltaprime-nD} for the hypercubic tiling this bound is as good as for the one-dimensional chain. This bound should hold also in general because is solely makes use of the normalization condition of the wave packet. We also checked in Figure \ref{fig:bounds-beta} whether this relation is satisfied for the labyrinth tiling and found that this is a significantly better lower bound than that given in Eq.\ \eref{equ:inequ-ketzmerick} for $w > w_{\rm th}$. However, in three dimensions there are still large deviations compared to $\beta$ for coupling parameters $w \to s$. This suggests that further studies are necessary in order to relate the width of the wave packet with the decay of its center.

The benefit of good lower bounds for $\beta$ is that the computation of the scaling exponents $\delta^\prime$, $D_2^\mu$, and $D_2$ is much less time consuming because for $\delta^\prime$ it is not necessary to average over all grid positions $\mathbf{r}$ and the spatial and spectral dimensions, $D_2$ and $D_2^\mu$, are even time-independent.

%******************************************************************************************************************************************
\section{Conclusion}\label{sec:conclusion}
%******************************************************************************************************************************************

In this paper we presented the scaling exponents of the participation ratio and the inverse participation numbers for large approximants of the separable Fibonacci tilings. The results showed that the wave functions of the systems are multifractals and that the strength of the multifractality increases with decreasing coupling parameter $w$ in agreement with the literature. By applying an RG approach we were able to show that the limit behaviour of the wave functions for $w \to 0$ is related to the structure of the Fibonacci chain. Up to now this observation was only based on numerical results.

Further, we presented numerical data for the quantum diffusion of electrons in these systems. Compared to previous results for the labyrinth tiling, we considered larger approximants and included an average over different initial positions of the wave packets, which allowed us to determine the scaling exponents with a higher precision. Our results also show that for the study of the quantum diffusion it is advisable to compute the return probability $P(t)$ instead of the temporal autocorrelation function because for an absolutely continuous energy spectrum the integral over the return probability converges and no further information can be obtained. In higher-dimensional systems $\delta \to 1$ usually corresponds to the existence of an absolutely continuous part in the energy spectrum rather than to the occurrence of ballistic transport.

In order to understand the diffusive properties of a system in detail it is also necessary to compute not only the width $d(t)$ of the wave packet but also the return probability $P(t)$. This can be easily understood from our results of the two tilings in three dimensions. While for the labyrinth tiling and the cubic Fibonacci tiling the scaling exponents $\beta$ hardly differ for the two models \cite{PhysRevB.2012.Thiem}, the scaling exponents $\delta^\prime$ are rather different for large coupling parameters $w$ (cf.\ Figure \ref{fig:paca-cmp}). In Figure \ref{fig:paca3D2} we have also seen that the wave packets spread in a quite different way for $w=s$. However, these differences are averaged out during the computations of the mean square displacement.

For the Fibonacci sequence we were also able to derive an analytical expression of the scaling exponent $\delta^\prime$ of the return probability $P(t)$ in the regime of strong quasiperiodic modulation, i.e., we found a relation between the quasiperiodic structure of the chain and the emerging electronic transport properties. Further, we show that the exponent $\delta^\prime$ can be used to define a better lower bound for the scaling exponent $\beta$ of the width of the wave packet for systems with absolutely continuous energy spectra. This result is not restricted to the quasiperiodic systems considered here.

\section*{References}

\newcommand{\noopsort}[1]{} \newcommand{\printfirst}[2]{#1}
\newcommand{\singleletter}[1]{#1} \newcommand{\switchargs}[2]{#2#1}

\end{document}